# Ligand additivity relationships enable efficient exploration of transition metal chemical space

Naveen Arunachalam[1,#], Stefan Gugler[1,#], Michael G. Taylor[1,#], Chenru Duan[1,2], Aditya Nandy[1,2], Jon Paul Janet[1], Ralf Meyer[1], Jonas Oldenstaedt[1], Daniel B. K. Chu[1], and Heather J. Kulik[1,*]

[1]*Department of Chemical Engineering, Massachusetts Institute of Technology, Cambridge, MA 02139, USA*

[2]*Department of Chemistry, Massachusetts Institute of Technology, Cambridge, MA 02139, USA*

#These authors contributed equally

ABSTRACT: To accelerate exploration of chemical space, it is necessary to identify the compounds that will provide the most additional information or value. A large-scale analysis of mononuclear octahedral transition metal complexes deposited in an experimental database confirms an under-representation of lower-symmetry complexes. From a set of around 1000 previously studied Fe(II) complexes, we show that the theoretical space of synthetically accessible complexes formed from the relatively small number of unique ligands is significantly (ca. 816k) larger. For the properties of these complexes, we validate the concept of ligand additivity by inferring heteroleptic properties from a stoichiometric combination of homoleptic complexes. An improved interpolation scheme that incorporates information about *cis* and *trans* isomer effects predicts the adiabatic spin-splitting energy to around 2 kcal/mol and the HOMO level to less than 0.2 eV. We demonstrate a multi-stage strategy to discover leads from the 816k Fe(II) complexes within a targeted property region. We carry out a coarse interpolation from homoleptic complexes that we refine over a subspace of ligands based on the likelihood of generating complexes with targeted properties. We validate our approach on 9 new binary and ternary complexes predicted to be in a targeted zone of discovery, suggesting opportunities for efficient transition metal complex discovery.

## 1. Introduction.

In recent years, virtual high-throughput screening (VHTS)[1-8] with first-principles density functional theory (DFT) and machine learning (ML) models[9-17] has greatly accelerated the discovery of new molecules and materials[18-23]. Nevertheless, the theoretical space of all possible compounds or materials is so large as to challenge even the most accelerated methods, with $10^{30}$ to $10^{60}$ theoretical drug-like molecules being enumerated[24-29] from a relatively small number of elements and atoms.[30, 31] Within the space of theoretical transition metal complexes, additional variables emerge, such as the metal identity, spin, and oxidation state, as well as denticity of the ligands.[4, 9] Indeed, significant analysis has been carried out by Fey and coworkers in understanding the role of privileged (e.g., phosphine) ligands in determining transition metal complex properties.[32-34] Jensen and coworkers have devised elegant strategies to explore the space of favored complexes, e.g., by adjusting the denticity during complex optimization carried out with efficient semi-empirical or force-field based scoring.[35-38] One key challenge for high-throughput screening with density functional theory of transition metal complex space is that the smallest non-trivial mononuclear octahedral complex consists of at least seven heavy atoms and nearly 100 electrons, also challenging the speed of conventional simulation techniques in comparison to readily computed data sets of closed-shell transition metal[39] and organic molecules[40, 41]. Further compounding the challenges of exploring transition metal chemical space are potential issues with convergence success or the presence of multi-reference character.[42-44] Thus, it is attractive to identify the minimal set of explicit first-principles calculations that can be used to build a model of the properties of the full data set.[45-47]

Recently, toward the goal of exploring a more diverse transition metal chemical space in comparison to complexes of frequently studied ligands, we devised a strategy for enumerating hypothetical, small (i.e., 1–2 heavy atoms per coordination site) ligands for mononuclear octahedral transition metal complexes[48]. These ligands sampled a diverse combination of coordinating atoms and their bonding environments[48], and only a small fraction were represented in prior databases of organic molecules.[25, 49, 50] We showed how incorporating these molecules could improve the fidelity of artificial neural network (ANN) models[48, 51] when applied to larger, realistic complexes present in the Cambridge Structural Database (CSD)[52]. That study was limited to the properties of homoleptic combinations of those ligands (i.e., all ligands equal), and therefore did not capture effects of mixing ligands that give rise to the compelling properties of many heteroleptic complexes and catalysts. Enumerating combinations of these ligands, however, would give rise to a combinatorial explosion, motivating strategies to understand which combinations are likely to be valuable or informative for a specific application.

Despite the challenge of combinatorial explosion, there are some established precedents of ligand additivity[53] that suggest that the properties of heteroleptic complexes can be inferred from combinations of homoleptic complexes[54]. For example, ligand additivity has been demonstrated in force field and DFT energetics[54] as well as DFT errors[55]. It has also been used in correction schemes, such as the DBLOC method.[56-59] We also recently exploited additivity to learn the degree of multireference character in a complex from the multireference character in its constituent ligands[60]. Additivity is also exploited heavily in fragmentation methods[61, 62] and in local correlated methods[63, 64]. In the present work, we carry out a survey of the symmetry classes and ligand diversity present in the CSD to confirm that the theoretical chemical space is orders of magnitude larger than the number that have been characterized. Motivated by the need to devise efficient but accurate methods for exploration of chemical space, we introduce improved interpolation schemes for heteroleptic compounds to incorporate *cis* and *trans* effects. Finally, we demonstrate how these approaches can be used for efficient but accurate discovery of transition metal complexes with targeted properties.

## 2. Computational Details.

### 2a. Curation from the Cambridge Structural Database.

A set of 85,575 mononuclear octahedral transition metal complexes was curated from the Cambridge Structural Database[52] (CSD) version 5.41 (Nov 2019). This procedure employed both the Conquest graphical interface to the CSD as well as the Python application programming interface, in all cases applied to the v5.41 data set with complexes from the November 2019 data set with both the March 2020 and May 2020 data updates (Supporting Information Text S1). For the complexes identified as octahedral, equatorial planes and axial positions were assigned based on prior reported rules.[65] To identify the symmetry of the ligands, unique ligands were identified by removing the metal atom to create independent molecular graphs for each ligand. Each ligand was identified as chemically unique within a given octahedral complex if it differed from all other ligands in the complex by: 1) heavy atom chemical symbols, 2) metal-connecting-atom element or 3) more than three hydrogens (Supporting Information Text S2). The symmetry of the complex was identified by distinguishing ligand denticity overall and in the equatorial plane along with the total number of unique ligands and whether ligands that were *trans* to each other were identical (Supporting Information Text S2). This led to a nomenclature for 66 ligand symmetry classes (Supporting Information Text S2).

To identify the set of unique ligands in each complex, a dummy atom with identical connectivity to the metal with atomic number of 0 was introduced to preserve the connectivity of the ligands to the metal without preserving metal identity. For this ligand and dummy atom combination, the atomic-number and bond-order weighted connectivity matrix determinant was calculated as described in Ref. [65]. We also computed the determinant of the atomic-number weighted bond-order weighted connectivity matrix where the off-diagonal elements, $Z_iZ_j$ ($i \neq j$), were set to the CSD-derived bond order for each ligand. Ligands with both distinct atomic-number-weighted connectivity matrix determinants and bond-order-weighted connectivity matrix determinants were identified as distinct ligands across monometallic transition metal complexes in the CSD. A second search was carried out by requiring that oxidation states and charges be assigned by the user along with no obvious disorder or missing hydrogen atoms in the structure, leading to 17,085 unique "computation-ready" complexes. Finally, we curated a subset of 1,202 Fe(II)-containing "computation-ready" complexes, based on the oxidation state reported by the user. From the ligands identified in this Fe(II) complex set, heteroleptic calculations from CSD ligands were carried out using a previously developed procedure[60] that enabled the assignment of the per-ligand charge.

**2b. Electronic Structure Calculations.**

DFT geometry optimizations were performed using a development version of TeraChem v1.9.[66, 67] The B3LYP[68-70] global hybrid functional was employed with the LANL2DZ[71] effective core potential for transition metals and the 6-31G* basis[72] for all other atoms. All transition metal complexes were studied with Fe(II) centers in low-spin singlet and high-spin quintet multiplicities. Singlet calculations were carried out in a spin-restricted formalism, while quintet calculations were unrestricted. Level shifting[73] was employed to aid self-consistent field convergence with the majority-spin and minority-spin virtual orbitals each shifted by 0.25 Ha. Geometry optimizations were carried out in the gas phase in translation rotation internal coordinates[74] using the BFGS algorithm. Default tolerances of $4.5 \times 10^{-4}$ hartree/bohr and $10^{-6}$ hartree were applied in the convergence criteria for the maximum gradient and energy difference between steps, respectively.

For the representative model complexes of $CH_3CN$, $H_2O$, CO, and $NH_3$, initial structures were generated with molSimplify[51, 75, 76], which uses OpenBabel[77, 78] as a backend. The same protocol was applied to generate homoleptic complexes of the 20 neutral ligands derived from monodentate-only, non-homoleptic Fe(II) complexes obtained from the CSD[52] after discarding one bulky ligand, $OP(Ph)_3$, that could not form a stable homoleptic structure for steric reasons. For the 36 homoleptic Fe(II) complexes in the CSD, the structures were directly extracted for subsequent geometry optimization. Heteroleptic complexes of the 12 representative ligands were also generated with molSimplify and optimized following the same procedure. A pickle file of the curated CSD database and structures and properties of all transition metal complexes studied with DFT is provided in the Supporting Information .zip file.

## 3. Results and Discussion.

### 3a. Symmetry Classes and Theoretical Complex Space.

The diversity present in the chemical space of transition metal complexes is derived from variability in the metal, its oxidation and spin state, as well as the chemistry of the coordinating ligands. Our experimental knowledge of this chemical space is unevenly distributed. We thus first examined the structures deposited in the CSD to uncover trends in the arrangement of ligands in previously characterized complexes (see Computational Details). From 85,575 mononuclear octahedral transition metal complexes of which 17,085 are identified as unique and computation-ready (e.g, have user-defined charges), the vast majority (95–98%) contain no more than three unique ligands (Figure 1 and Supporting Information Table S1). In fact, 28% of all unique computation-ready complexes are homoleptic, and a majority (76%) contain no more than two ligand types (Figure 1 and Supporting Information Table S1). Because metal identity and ligand diversity are expected to be coupled, we also evaluated statistics on a subset of 1,202 unique Fe(II) complexes and confirm that the preference for complexes with no more than two ligands is preserved and even strengthened (Figure 1 and Supporting Information Table S1).

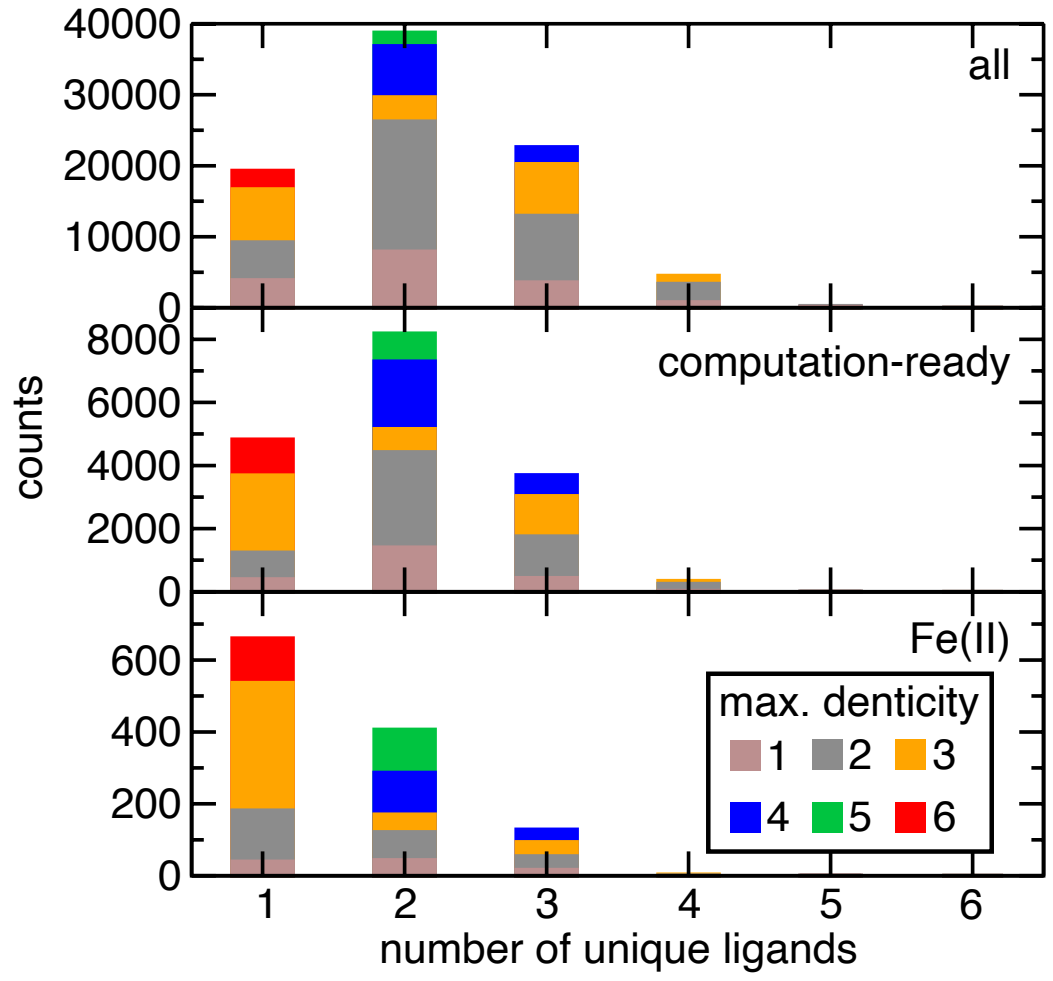

**Figure 1.** Stacked, unnormalized histogram of the number of complexes in the CSD grouped by number of unique ligand types and by the highest denticity of ligands in the complex (monodentate in brown, bidentate in gray, tridentate in orange, tetradentate in blue, pentadentate in green, and hexadentate in red, as indicated in inset legend). These counts are shown for all complexes (top), unique and computation-ready complexes (middle), and the Fe(II) unique, computation-ready subset (bottom).

To identify how the structures sampled in the CSD compare to the theoretical space of all hypothetical complexes, we enumerate the overall pool of theoretical complexes and the number in each symmetry class to compare to the most frequently characterized symmetry classes in the CSD. We applied Pólya's enumeration theorem to octahedral coordination geometries to obtain all possible symmetry classes from the cycle index (i.e., theoretical sum) of a symmetry group.[79-81] The cycle index for all possible octahedral complexes is

$$Z(\mathrm{O_h}) = \frac{1}{48}\left(f_1^6 + 8f_3^2 + 7f_2^3 + 6f_1^2 f_4 + 9f_1^2 f_2^2 + 8f_6 + 3f_1^4 f_2 + 6f_4 f_2\right) \quad (1)$$

where the subscript, *j,* on *f* indicates the number of unique sites (e.g., 2 for $M(L_1)_5(L_2)_1$) up to a theoretical maximum of 6. Each individual term in the cycle index is computed as:

$$f_j = x_1^j + \cdots + x_N^j \quad (2)$$

where $N$ is the total number of unique ligands that can occupy the sites, and each $x_i$ corresponds to one unique ligand set choice.

For a worked example of $N$ = 12 unique ligands, we determine the theoretical number of complexes, the cardinality, as

$$\frac{N!}{(N-M)\prod_{k=1}^{6}(\sum_{j=1}^{M}\mathbb{I}_{|L_j|=k})!} \tag{3}$$

where its respective symmetry class $C = \{L_1, \ldots, L_M\}$ is built from $M$ distinguishable sites, $j$, and $N$ are the ligand index and number of ligands as before. The indicator function, $\mathbb{I}_{|L_j|=k}$, where e.g., $|L_1|$ would be 6 for $(L_1)_6$, counts how many occurrences of each substituent combination exist. For example, in $(L_1)_4(L_2)_2$ there is one occurrence of a 4-substitution site and 1 occurrence of a 2-substitution site, whereas in $(L_1)_2(L_2)_2(L_3)_2$, there are 3 occurrences of a 2-substitution site. For low-symmetry cases, this generalized equation simplifies significantly (Table 1).

**Table 1.** The number of theoretical complexes for each octahedral symmetry class considered in this work as well as the full octahedral space for an example single metal/oxidation/spin state ($m$ = 1) with an $N$ = 12 ligand pool. The configurations and isomers indicate the number of ways unique ligands can be arranged, and the cardinality indicates how many theoretical complexes can be enumerated.

| **Name** | **Configuration** | **Isomers** | **Cardinality** | **Complexes** |
|---|---|---|---|---|
| HO | $x^6$ | 1 | $N$ | 12 |
| 5+1 | $x^1x^5$ | 1 | $N(N\text{-}1)$ | 132 |
| TS/CS | $x^4x^2$ | 2 | $N(N\text{-}1)$ | 264 |
| FS/MS | $x^3x^3$ | 2 | $N(N\text{-}1)/2!$ | 132 |
| CA/TA | $x^4x^1x^1$ | 2 | $N(N\text{-}1)(N\text{-}2)/2!$ | 1,320 |
| FA/MAC/MAT | $x^3x^2x^1$ | 6 | $N(N\text{-}1)(N\text{-}2)/2!$ | 3,960 |
| EA/DCS/DTS | $x^2x^2x^2$ | 5 | $N(N\text{-}1)(N\text{-}2)/3!$ | 1,100 |
| Up to two ligands | | | | 540 |

| | | |
|---|---|---|
| Up to three ligands | | 6,920 |
| Full | $1/48\,(N^6 + 3N^5 + 9N^4 + 13N^3 + 14N^2 + 8N)$ | 82,160 |

For the trivial case of homoleptic (HO) complexes, $N$ ligands produce $N$ complexes (i.e., 12 for $N$ = 12, Table 1). For up to two unique ligand types, the five unique symmetry classes consist of monoheteroleptic (5+1) $M(L_1)_5(L_2)_1$ complexes, *trans* symmetric (TS) or *cis* symmetric (CS) $M(L_1)_4(L_2)_2$ complexes, as well as *fac* symmetric (FS) or *mer* symmetric (MS) $M(L_1)_3(L_2)_3$ complexes (Figure 2 and Table 1). Both 5+1 and CS/TS complexes each form $N(N\text{-}1)$ complexes for $N$ ligands (i.e., 132 each for $N$ = 12, Table 1). The degeneracy of the stoichiometry in FS and MS complexes gives rise to $N(N\text{-}1)/2!$ complexes (i.e., 66 each for $N$ = 12, Table 1). Thus, from $N$ = 12 ligands, a total of 540 complexes may be formed with up to two unique ligand types.

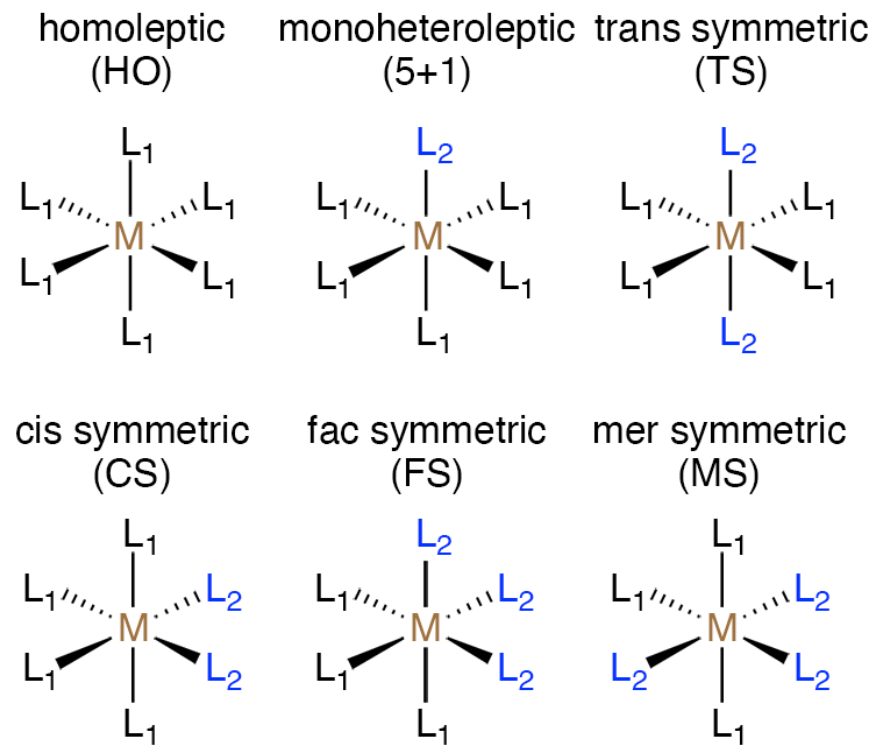

**Figure 2.** Symmetry classes for transition metal complexes with up to two unique ligands, $L_1$ and $L_2$ from left to right and top to bottom: homoleptic (HO) $M(L_1)_6$, monoheteroleptic (5+1) $M(L_1)_5L_2$, *trans* symmetric (TS) $M(L_1)_4(L_2)_2$, *cis* symmetric (CS) $M(L_1)_4(L_2)_2$, *fac* symmetric (FS) $M(L_1)_3(L_2)_3$, and *mer* symmetric (MS) $M(L_1)_3(L_2)_3$. For each pair of ligands, a total of two homoleptic and eight two-ligand isomers can be obtained because the 5+1, *trans* symmetric, and *cis* symmetric complexes are unique if the stoichiometry of $L_1$ and $L_2$ are swapped.

Expanding to up to three unique ligand types introduces eight additional symmetry classes (Figure 3). These include $M(L_1)_4L_2L_3$ *cis* asymmetric (CA) and *trans* asymmetric (TA) complexes, the three types of $M(L_1)_2(L_2)_2(L_3)_2$ configurations in equatorial asymmetric (EA), double *cis* symmetric (DCS), or double *trans* symmetric (DTS) symmetries, and three $M(L_1)_2(L_2)_3L_3$ in *fac* asymmetric (FA), *mer* asymmetric *trans* (MAT), or *mer* asymmetric *cis* (MAC) symmetries (Figure 3). For the EA complexes, there are 3 occurrences of a 2-substitution site that fulfill the EA definition. This combines with the DCS and DTS isomers to form a total of 5 isomers with $N(N\text{-}1)(N\text{-}2)/3!$ possible combinations of ligands (i.e., 1100 for $N = 12$, Table 1). The less degenerate CA/TA complexes form a total of $N(N\text{-}1)(N\text{-}2)/2!$ complexes for each of the two isomers (i.e., 1320 for $N = 12$, Table 1). Similarly, FA/MAC/MAT complexes can each form $N(N\text{-}1)(N\text{-}2)/2!$ complexes in two isomers each (i.e., 3960 for $N = 12$, Table 1).

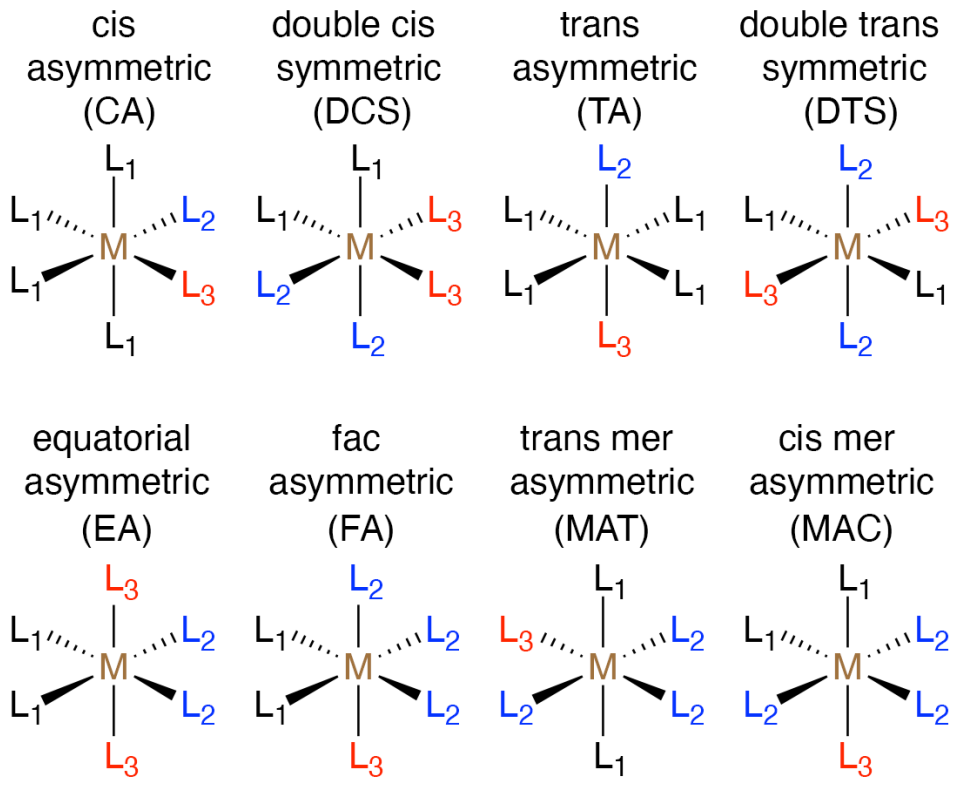

**Figure 3.** Symmetry classes for transition metal complexes with three unique ligands, $L_1$,$L_2$, and $L_3$, from left to right and top to bottom: *cis* asymmetric (CA) $M(L_1)_4L_2L_3$, double *cis* symmetric (DCS) $M(L_1)_2(L_2)_2(L_3)_2$, *trans* asymmetric (TA) $M(L_1)_4L_2L_3$, double *trans* symmetric (DTS) $M(L_1)_2(L_2)_2(L_3)_2$, equatorial asymmetric (EA) $M(L_1)_2(L_2)_2(L_3)_2$, *fac* asymmetric (FA) $M(L_1)_2(L_2)_3L_3$, *trans mer* asymmetric (MAT) $M(L_1)_2(L_2)_3L_3$, and *cis mer* asymmetric (MAC) $M(L_1)_2(L_2)_3L_3$. A total of 29 complexes can be obtained for any combination of three ligands due to additional isomers of the equatorial asymmetric type as well as those for which the stoichiometry of each ligand type is not equal.

In total, 6,920 complexes can be formed from $N = 12$ ligands for the three symmetry classes considered here. For the same ligand pool, there is a much larger set of 82,160 theoretical complexes that could be created from a greater number of unique ligands. This analysis does not consider cases where the ligand chemistry prevents formation of a complex, e.g., only monodentate and pentadentate ligands can form the 5+1 symmetry class, whereas monodentate, bidentate, tridentate, or hexadentate ligands can form HO complexes.

Returning to the diversity observed in complexes deposited in the CSD, we can qualitatively observe that relatively little of the theoretical space has been sampled. As we have shown for a representative example, for any set of unique ligands a much larger theoretical number of binary and ternary complexes can be formed in comparison to homoleptic complexes. Nevertheless, there are far fewer binary and especially ternary complexes in the CSD (Supporting Information Table S1). Within the binary complexes, 5+1 and TS complexes are overrepresented in comparison to those for FS, MS, or CS based on our theoretical enumeration (Figure 4 and Supporting Information Table S2). All ternary complexes are underrepresented, but those with equal stoichiometry (i.e., EA, DCS, or DTS) have very few examples in the CSD (Figure 4 and Supporting Information Table S2). If we simplify our analysis by focusing only on Fe(II) complexes, the same trends hold, although Fe(II) complexes have an even greater relative number of 5+1 and TS complexes and lower number of CS complexes (Figure 4 and Supporting Information Table S2).

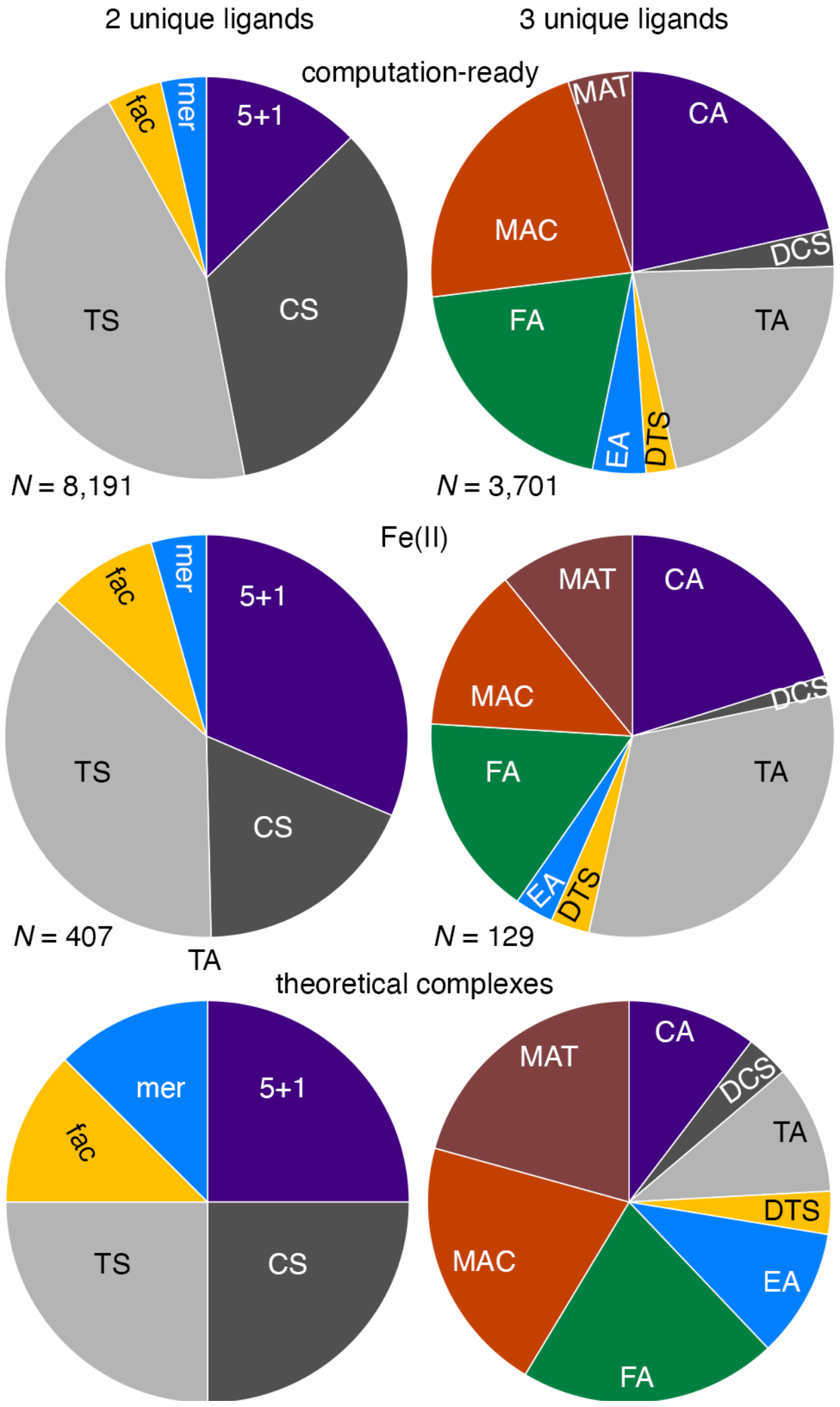

**Figure 4.** Percent of all unique mononuclear octahedral transition metal complexes in the CSD with user-defined charges (top, computation-ready) as well as the Fe(II) subset (middle, Fe(II)), grouped by symmetry class for cases with two unique ligands (left: 5+1, *cis* symmetric, CS, *trans* symmetric, TS, *fac*, or *mer*) or three unique ligands (right: *cis* asymmetric, CA, double *cis* symmetric, DCS, *trans* asymmetric, TA, double *trans* symmetric, DTS, equatorial asymmetric, EA, *fac* asymmetric, FA, *cis mer* asymmetric, MAC, or *trans mer* asymmetric, MAT). The ratio of symmetry classes for the theoretical complexes from enumeration are shown at bottom for comparison.

An additional factor in analysis of symmetry classes is the extent to which ligand denticity plays a role. While enumeration is straightforward for monodentate ligands, higher-denticity ligands (e.g., pentadentates) may only be compatible with some of the symmetry classes. Indeed, over both the computation-ready CSD set and the Fe(II) subset, a significant number of complexes consist of higher-denticity ligands that would be incompatible with full enumeration (Figure 1 and Supporting Information Tables S3–S4). Some multidentate ligands are also restricted in which symmetry classes they can form due to rigidity, while other more flexible ligands are less restricted. Nevertheless, we can still conclude there is vastly higher sampling of homoleptic structures. For example, tridentate ligands can be present in a homoleptic complex as well as any binary FS/MS or ternary TA/FA/MAC/MAT complex. Nearly an order-of-magnitude more unique tridentate ligands have been characterized in homoleptic Fe(II) complexes in comparison to either the binary or ternary cases (Supporting Information Table S4). Similar trends hold for all unique computation-ready complexes (Supporting Information Table S3). Overall, the number of unique ligands for binary or ternary complexes is still lower than the number of unique complexes, but the gap is smaller than could be expected from enumeration alone (Supporting Information Tables S3–S4).

To simplify a quantitative comparison between the theoretical space and the enumerated space, we focus on Fe(II) complexes with only monodentate ligands in any of the fourteen symmetry classes considered. For this set, we identify 40 unique monodentate ligands present in 40 HO complexes (Supporting Information Table S5). There are an additional 48 ligands present in binary or ternary complexes not observed in homoleptics for which we can confidently assign a charge to the ligand (Supporting Information Table S5). While the majority of HO complex monodentate ligands were neutral in charge, over half of the additional binary or ternary ligands have a non-zero charge (Supporting Information Table S6). Thus, although we identify 88 unique monodentate ligands in previously synthesized Fe(II) complexes, ligands not sampled in HO complexes may be incompatible with the HO symmetry class if they give rise to high overall complex charges.

Taking the set of 88 ligands from any HO, binary, or ternary Fe(II) complex as the theoretical space for which compatibility across symmetry classes should be maximal, we then quantified the theoretical versus actual coverage of Fe(II) complex chemical space. A significant number (39%) of all HO complexes have been characterized. In comparison, the binary symmetry classes have not been as well explored, with TS complexes the highest at 0.3% (i.e., 26 of 7,656 theoretical complexes and Supporting Information Table S6). The number of theoretical ternary complexes grows rapidly with this ligand pool, ranging from 109,736 theoretical DTS/DCS complexes to 329,208 FA/MAC/MAT/CA/TA complexes (Supporting Information Table S6). In total, the 1,202 Fe(II) complexes represent a tiny fraction of the theoretical 3,213,056 homoleptic, binary, or ternary complexes that could form from 88 experimentally synthesized ligands. Of the 88 ligands, we exclude one (i.e., $OP(Ph)_3$) from further analysis due to its large bulk that prevents building a homoleptic complex that remains intact after geometry optimization. Even if we restrict ourselves to the 56 ligands that are neutral, closed-shell singlets, and amenable to homoleptic complex construction, the theoretical space of homoleptic, binary, or ternary complexes is still large (i.e., 816,256 complexes). Thus, we conclude that efficient strategies to infer heteroleptic properties from homoleptic properties are necessary to "fill in" the remainder of this unexplored space.

**3b. Ligand Additivity for Interpolating Properties of Transition Metal Complexes.**

We next aimed to determine the extent to which the properties of lower-symmetry heteroleptic transition metal complexes could be inferred from those of higher-symmetry complexes. We constructed complexes with two to three unique ligand types from small sets of ligands that spanned a large range of ligand field strengths: weak-field water, strong-field carbonyl, and either strong-field methylisocyanide or weak-field ammonia. For calculation of the adiabatic high-spin (e.g., quintet Fe(II)) to low-spin (e.g., singlet Fe(II)) splitting, $\Delta E_{\text{H-L}}$, it can be expected that the weak-field ligands will lead to homoleptic complexes that favor high-spin states, whereas strong-field ligands will make homoleptic complexes that favor low-spin states. Thus, heteroleptic combinations of these ligands are expected to reside between the two limits. One simple way to obtain estimates of the spin splitting of the heteroleptic complexes (e.g., with up to three unique ligand types) is to take a weighted average of the spin splitting of the parent homoleptic complexes:

$$E(M(L_1)_x(L_2)_y(L_3)_{6-x-y}) = \frac{x}{6}E(M(L_1)_6) + \frac{y}{6}E(M(L_2)_6) + \frac{6-x-y}{6}E(M(L_3)_6) \quad (4)$$

Indeed, we observe that heteroleptics reside between the homoleptic limits for spin splitting, and the linear averaging roughly holds for complexes with methyl isocyanide ($CH_3CN$), $H_2O$, and CO ligands (Figure 5 and Supporting Information Table S7). Similar observations can be made on combinations with $H_2O$, CO, and $NH_3$ (Supporting Information Figure S1). Nevertheless, there are significant outliers in the interpolated versus actual $\Delta E_{\text{H-L}}$, which are particularly evident when comparing heteroleptic complexes with the same stoichiometry, and therefore the same prediction from a simple linear model, but distinct ligand symmetry (Figure 5). For example, the CS complex Fe(II)($H_2O$)$_4$(CO)$_2$ $\Delta E_{\text{H-L}}$ is predicted accurately (predicted: -7.9 kcal/mol vs calculated: -9.9 kcal/mol) from homoleptic interpolation (Figure 5 and Supporting Information Table S8). The same prediction significantly overestimates the TS complex with the same stoichiometry (predicted: -7.9 kcal/mol vs calculated: -19.3 kcal/mol), which instead behaves much more similarly to the 5+1 complex, as observed in prior work[82] (Figure 5). Overall, for this combination of ligands, CS or FS complexes that have one minority ligand in the axial position and another in the equatorial plane are much better predicted than the equivalent MS or TS complexes (Figure 5). For the case of CO, $H_2O$, and $NH_3$, mixing between weak-field $NH_3$ and $H_2O$ is relatively accurately predicted from homoleptic averaging, whereas for $NH_3$ and CO, it is the MS and TS complexes that are more accurately predicted than the FS or CS counterparts (Supporting Information Figure S2 and Tables S8–S9).

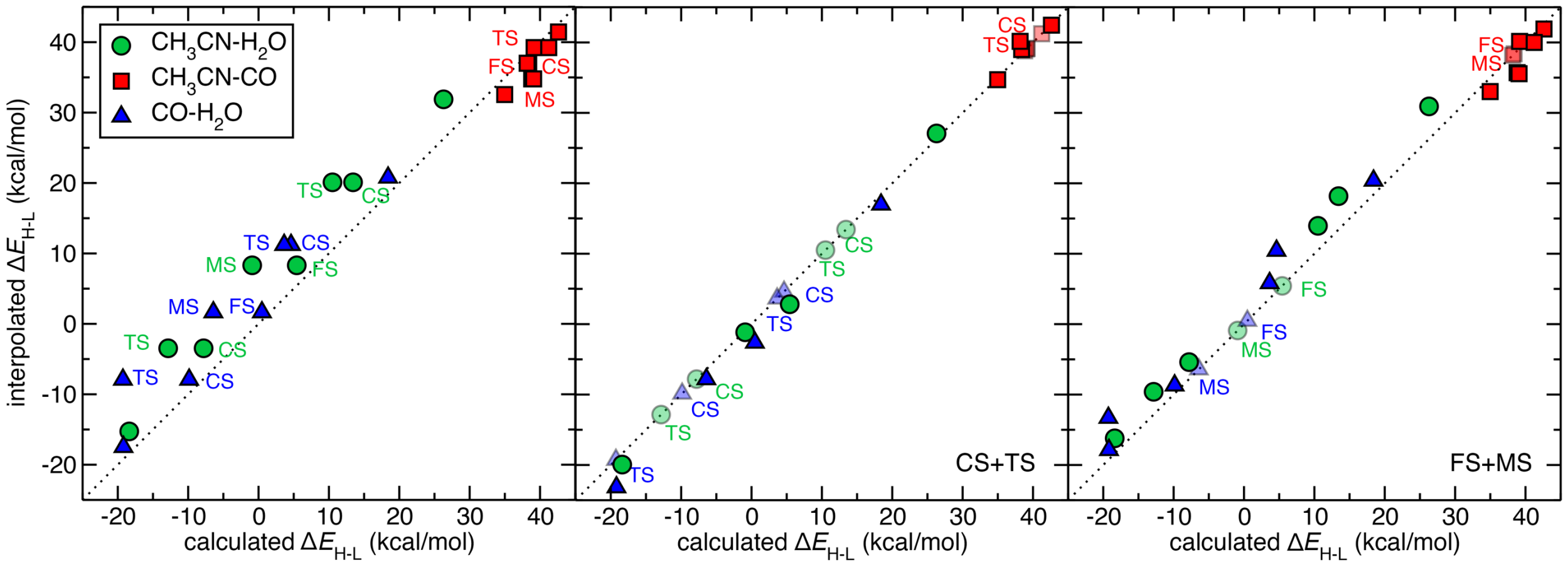

**Figure 5.** Calculated vs. linearly interpolated $\Delta E_{\text{H-L}}$ (kcal/mol) for Fe(II) complexes with pairs of any of the three ligands: $CH_3CN$, $H_2O$, and CO. From left to right: interpolation between homoleptic complexes (HO only), interpolation using homoleptic complexes as well as CS and TS complex energies (CS+TS), or interpolation using homoleptic complexes as well as FS and MS complex energies (FS+MS). Points are colored according to the pair of ligands they correspond to: $CH_3CN$-$H_2O$ (green circles), $CH_3CN$-CO (red squares), and CO-$H_2O$ (blue triangles), as indicated in inset legend. Key isomers are annotated. Points provided for the fit are translucent, whereas the remaining points are opaque. In all panes, a black dotted parity line is shown.

Because adiabatic spin splitting involves geometry optimizations in two distinct spin states, we also evaluated HOMO levels of each singlet complex as a property that depends only on a single geometry. Overall, interpolation of HOMO energies of heteroleptic complexes of $CH_3CN$, CO, and $H_2O$ from homoleptic complexes reproduces trends between the homoleptic limits (Figure 6 and Supporting Information Tables S7, S10–S11). As with spin splitting, there are key differences for complexes with identical stoichiometry that cannot be captured by interpolation from homoleptic complexes alone (Figure 6). Interestingly, for the same complexes for which the CS complex had a higher (i.e., more off-parity) $\Delta E_{\text{H-L}}$ and the TS complex was more like the equivalent 5+1 complex, we find more varied results for the HOMO level, with some cases occurring where the TS and FS complexes are less accurately predicted (Figure 6 and Supporting

Information Table S10). Nevertheless, over the full range of data, the outliers in the HOMO level prediction are more modest than was observed for $\Delta E_{H\text{-}L}$. These trends also hold for the complexes with $NH_3$, $H_2O$, and CO (Supporting Information Figure S2 and Tables S10–S11).

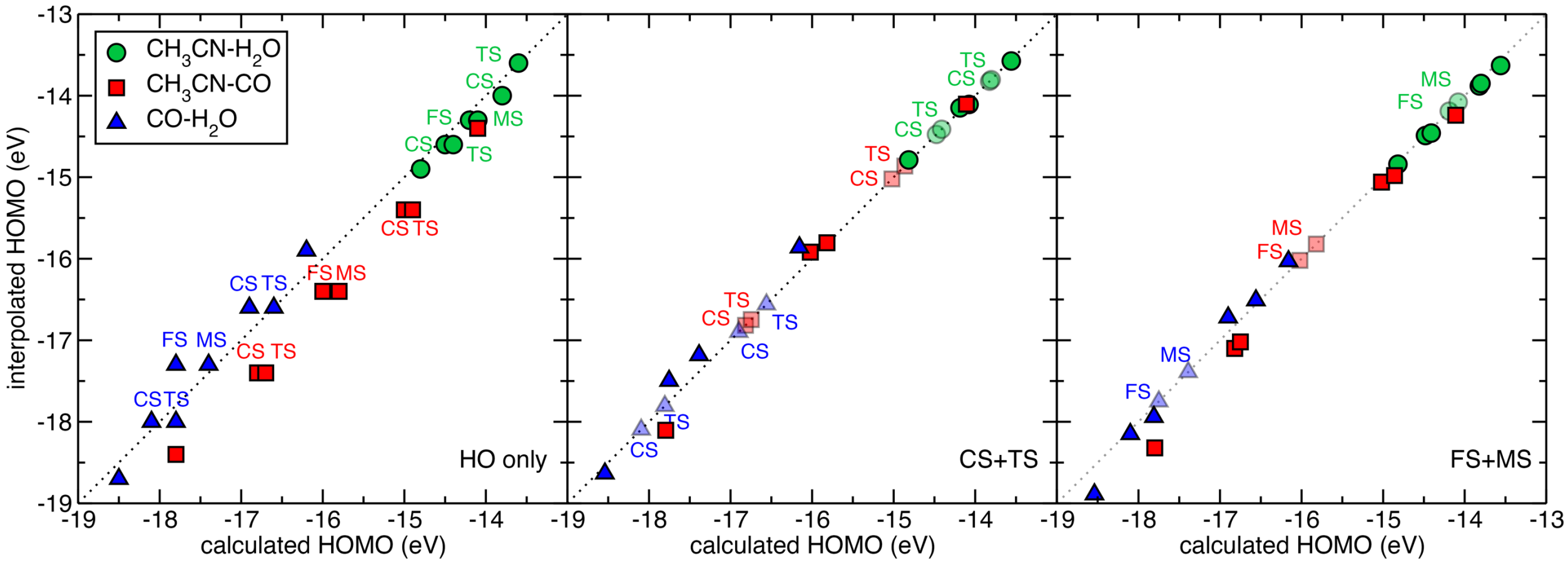

**Figure 6.** Calculated vs. linearly interpolated HOMO level (eV) for singlet Fe(II) complexes with pairs of any of the three ligands: $CH_3CN$, $H_2O$, and CO. From left to right: interpolation between homoleptic complexes (HO only), interpolation using homoleptic complexes as well as CS and TS complex energies (CS+TS), or interpolation using homoleptic complexes as well as FS and MS complex energies (FS+MS). Points are colored according to the pair of ligands they correspond to: $CH_3CN$-$H_2O$ (green circles), $CH_3CN$-CO (red squares), and CO-$H_2O$ (blue triangles), as indicated in inset legend. Key isomers are annotated. Points provided for the fit are translucent, whereas the remaining points are opaque. In all panes, a black dotted parity line is shown.

Given the differences between CS and TS complexes despite having identical stoichiometry, we next identified strategies for improving the interpolation. First, we employed CS and TS complexes along with homoleptic complexes to predict the spin-splitting energetics of the other binary heteroleptic complexes as follows:

$$E(5+1) = \frac{1}{2}(E(HO) + E(TS)) \quad (5)$$

where $E(5+1)$ is the interpolated energy of the $M(L_1)_5L_2$ complex from the $M(L_1)_6$ HO energy and the $M(L_1)_4(L_2)_2$ TS energy.

Similarly, we estimate the FS and MS energies as:

$$E(FS) = \frac{1}{2}(E(CS\ M(L_1)_4(L_2)_2) + E(CS\ M(L_2)_4(L_1)_2)) \quad (6)$$

$$E(MS) = \frac{1}{2}(E(TS\ M(L_1)_4(L_2)_2) + E(TS\ M(L_2)_4(L_1)_2)) \quad (7)$$

In practice this corresponds to computing six energies and interpolating four remaining energies for each pair of ligands for a computational savings of 40%. Indeed, we observe reduced mean absolute error (MAE) over the remaining points that are interpolated (2.4 kcal/mol vs 5.1 kcal/mol) in comparison to the HO-only interpolation (Figure 5 and Supporting Information Tables S8–S9). In particular, FS complexes of CO and $H_2O$ are now correctly predicted to be much more low-spin-directing than the MS complex of the same stoichiometry (Figure 5). The HOMO levels for these and other complexes are also improved (Supporting Information Tables S10–S11). For example, the modified interpolation is able to capture the fact that MS Fe(II)$(CO)_4(H_2O)_2$ has a shallower HOMO level than its FS counterpart (Figure 6). Nevertheless, not all points are uniformly improved by this interpolation. For the case of $NH_3$, CO, and $H_2O$ where the interpolation already performed well, some points such as the $\Delta E_{\text{H-L}}$ for 5+1 Fe(II)$(CO)_5NH_3$ are slightly worsened in the modified interpolation (Supporting Information Figure S1). For the same complex, the HOMO level is equivalently predicted by both interpolation schemes, and most HOMO level estimates are improved (Supporting Information Figure S2). Overall, errors are on average significantly lower for all properties and sets of ligands considered when the modified interpolation expressions are employed.

Although the CS/TS-derived interpolation schemes greatly reduce errors in estimating the energetics of heteroleptic complexes, they still require significant computational overhead. Thus, we next aimed to identify if FS and MS complexes, which contain three ligands *cis* to each other or two ligands *cis* and two sets of ligands *trans*, respectively, could be used instead in the interpolation (see Figure 2). If the FS and MS complexes impart sufficient information, using them in an interpolation scheme along with homoleptic complex properties corresponds to evaluating four properties (e.g., energies) to predict six properties for a computational savings of 60%. In this interpolation scheme, we estimated the complex properties from FS and MS complexes as follows:

$$E(5+1) = \frac{2}{3}E(HO) + \frac{1}{3}E(FS) \tag{8}$$

where $E$(5+1) is the interpolated energy of the $M(L_1)_5L_1$ complex from the $M(L_1)_6$ HO energy and the $M(L_1)_3(L_2)_3$ FS energy. Similarly, we obtain expressions for the CS and TS complexes as:

$$E(CS) = \frac{1}{3}E(HO) + \frac{2}{3}E(FS\, M(L_1)_3(L_2)_3) \tag{9}$$

$$E(TS) = \frac{1}{3}E(HO) + \frac{2}{3}E(MS\, M(L_1)_3(L_2)_3) \tag{10}$$

Indeed, using this approach we achieve errors only slightly larger than that for the CS/TS averaging scheme, with the added benefit of requiring fewer energies to obtain the same fidelity of the interpolation (Figure 5). For example, the higher spin-splitting energy of CS complexes relative to TS complexes is captured here because it can be directly derived from the strength of the high-spin directing character of the FS complex relative to that of the MS complex (Figure 5). Mixing of the weak-field $NH_3$ and $H_2O$ ligands, which had slightly worsened with CS/TS interpolation, is also significantly improved in this scheme, and the other sets of complexes are of comparable accuracy (Supporting Information Figure S1 and Tables S8–S9). While for the HOMO level, CS/TS interpolation tended to underestimate the HOMO level, FS/MS interpolation slightly overestimates HOMO levels, but errors are much smaller than for the HO-only interpolation (Figure 6 and Supporting Information Tables S10–S11). For the set of ligands including ammonia, almost all points are predicted to comparable or slightly improved values (Supporting Information Figure S2). Thus, HO-only interpolation provides a highly efficient scheme for predicting heteroleptic transition metal complexes, but the best trade-off in accuracy and computational cost for interpolation is likely achieved through estimating properties using information from FS and MS complexes.

We next investigated whether we could generalize our observations to heteroleptics with three unique ligand types (Figure 3). This extension is motivated by the fact that 96% of all mononuclear transition metal complexes in the CSD contain no more than three ligand types, and over 99% of Fe(II) complexes contain three or fewer ligand types (Figure 1 and Supporting Information Table S1). Expansion to three ligand types introduces 29 complex energies that need computation for any set of three ligands. Homoleptic-only averaging performs poorly here for complexes that are mixtures of $CH_3CN$, CO, and $H_2O$, with a large difference between CA and TA complexes of $H_2O$ and CO being treated completely equivalently in this scheme (Figure 7). Similar differences in TA and CA HOMO levels are also missed in this averaging scheme (Figure 8). Generally, the CA complex spin-splitting energies are better predicted by the homoleptic averaging than the TA are for both ternary complexes with $CH_3CN$ and with $NH_3$ (Figure 7 and Supporting Information Figure S3 and Tables S12–S13). For the HOMO level, results are more varied, with the HOMO energies of the $CH_3CN$ ternary complexes being underestimated while those of $NH_3$ ternary complexes are overestimated (Figure 8 and Supporting Information Figure S4 and Tables S14–S15).

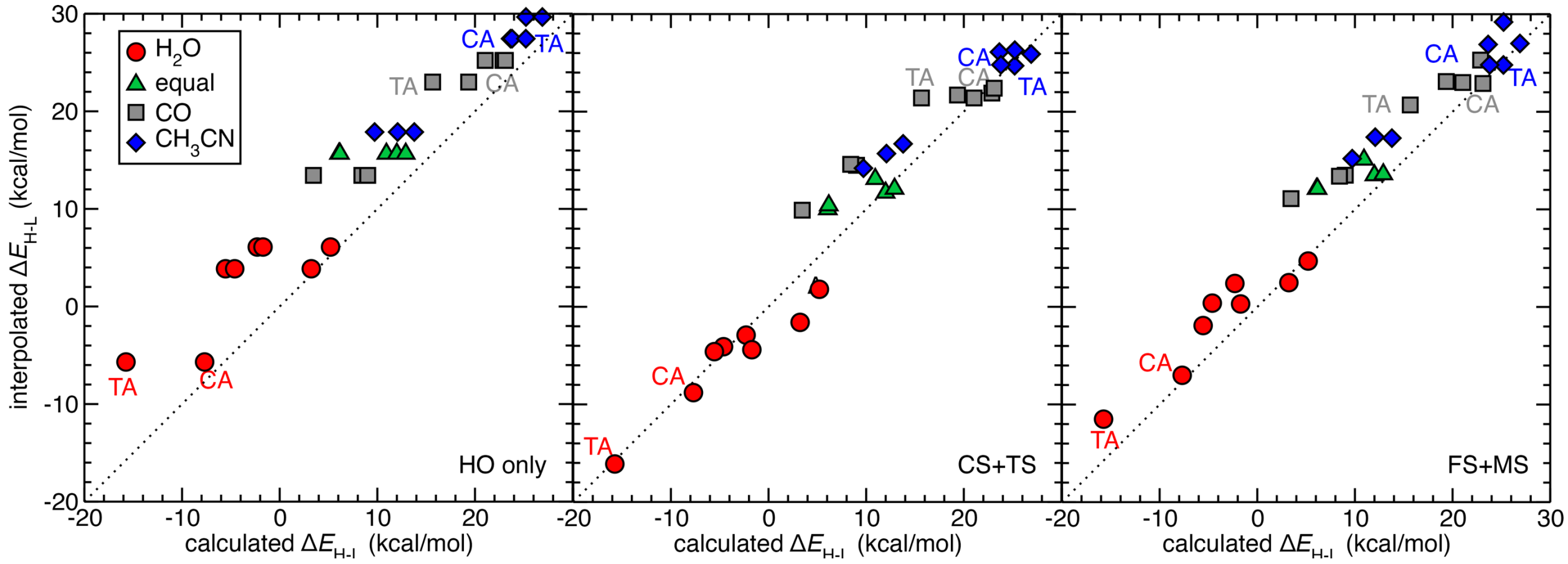

**Figure 7.** Calculated vs. linearly interpolated $\Delta E_{\text{H-L}}$ (kcal/mol) for Fe(II) complexes with at least one each of three ligands: $CH_3CN$, $H_2O$, and CO. From left to right: interpolation from homoleptic complexes (HO only), interpolation using homoleptic complexes as well as CS and TS complex energies derived from pairs of ligands (CS+TS), or interpolation using homoleptic complexes as well as FS and MS complex energies derived from pairs of ligands (FS+MS). Points are colored according to the ligand with the highest stoichiometric coefficient: $H_2O$ (red circles), CO (gray squares), and $CH_3CN$ (blue diamonds), or equal weight of all ligands (Green triangles), as indicated in inset legend. Key isomers are annotated. In all panes, a black dotted parity line is shown.

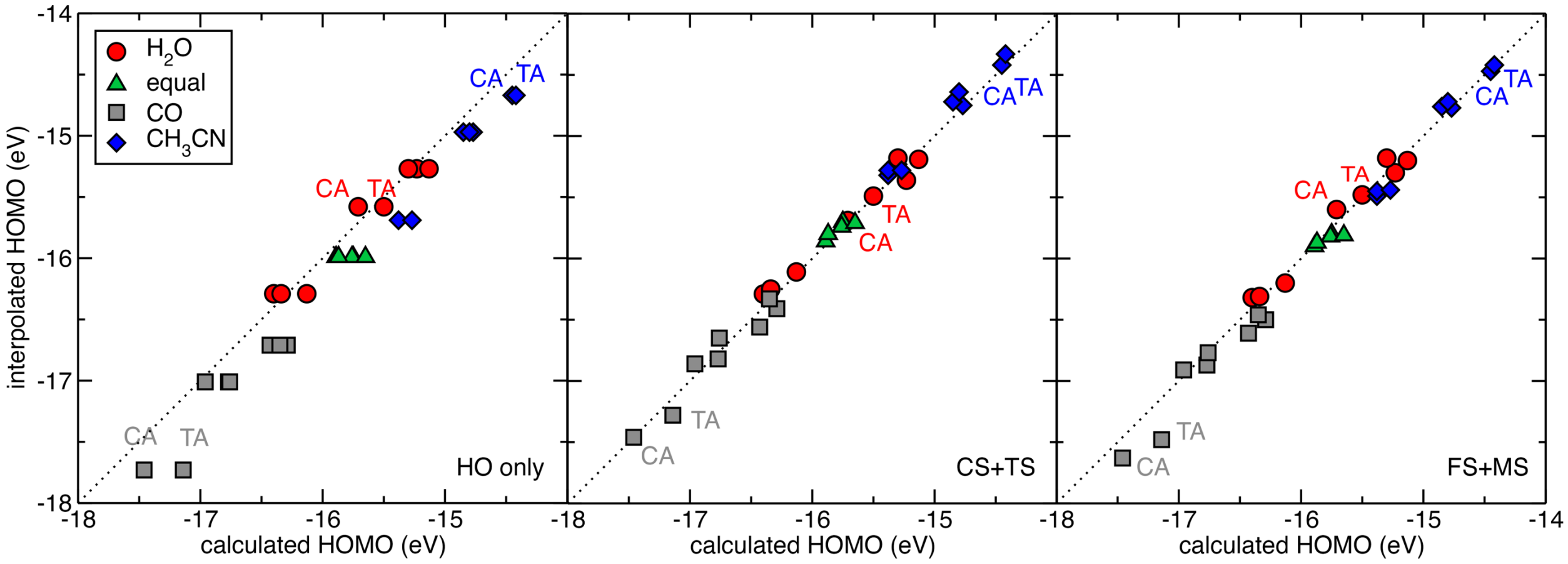

**Figure 8.** Calculated vs. linearly interpolated HOMO level (eV) for singlet Fe(II) complexes with at least one each of three ligands: $CH_3CN$, $H_2O$, and CO. From left to right: interpolation from homoleptic complexes (HO only), interpolation using homoleptic complexes as well as CS and TS complex energies derived from pairs of ligands (CS+TS), or interpolation using homoleptic complexes as well as FS and MS complex energies derived from pairs of ligands (FS+MS). Points are colored according to the ligand with the highest stoichiometric coefficient: $H_2O$ (red circles), CO (gray squares), and $CH_3CN$ (blue diamonds), or equal weight of all ligands (green triangles), as indicated in inset legend. Key isomers are annotated. In all panes, a black dotted parity line is shown.

Thus, we next investigated interpolation schemes that reincorporated CS/TS or FS/MS complex energies of the binary heteroleptics. Interpolation of these 29 energies from 12 CS/TS and 3 HO energies or 6 FS/MS and 3 HO energies would still represent significant computational savings. The binary complex energies can also be reused for ternary complexes that share a pair of ligand types, as is the case in the two worked examples presented here. Next, we obtained expressions for all eight symmetry classes of heteroleptics with three ligand types from energies derived from only either FS/MS or CS/TS complexes. Specifically, the FS/MS-interpolated expressions for these heteroleptics are as follows:

$$E(CA\ M(L_1)_4(L_2)_1(L_3)_1) = \frac{1}{3}(E(HO\ M(L_1)_6) + E(FS\ M(L_1)_3(L_2)_3) + E(FS\ M(L_1)_3(L_3)_3)) \quad (11)$$

$$E(TA\ M(L_1)_4(L_2)_1(L_3)_1) = \frac{1}{3}(E(HO\ M(L_1)_6) + E(MS\ M(L_1)_3(L_2)_3) + E(MS\ M(L_1)_3(L_3)_3)) \quad (12)$$

$$E(FA\ M(L_1)_3(L_2)_2(L_3)_1) = \frac{1}{3}(\frac{1}{2}E(HO\ M(L_1)_6) + \frac{1}{2}E(HO\ M(L_2)_6) + E(FS\ M(L_1)_3(L_2)_3) + E(FS\ M(L_1)_3(L_3)_3)) \quad (13)$$

$$E(MAC\ M(L_1)_3(L_2)_2(L_3)_1) = \frac{1}{3}(\frac{1}{2}E(HO\ M(L_1)_6) + \frac{1}{2}E(HO\ M(L_2)_6) + E(FS\ M(L_1)_3(L_2)_3) + E(MS\ M(L_1)_3(L_3)_3)) \quad (14)$$

$$E(MAT\ M(L_1)_3(L_2)_2(L_3)_1) = \frac{1}{3}(\frac{1}{2}E(HO\ M(L_1)_6) + \frac{1}{2}E(HO\ M(L_2)_6) + E(MS\ M(L_1)_3(L_2)_3) + E(MS\ M(L_1)_3(L_3)_3)) \quad (15)$$

$$E(DCS\ M(L_1)_2(L_2)_2(L_3)_2) = \frac{1}{3}(\frac{1}{3}E(HO\ M(L_1)_6) + \frac{1}{3}E(HO\ M(L_2)_6) + \frac{1}{3}E(HO\ M(L_3)_6) + \frac{2}{3}E(FS\ M(L_1)_3(L_2)_3) + \frac{2}{3}E(FS\ M(L_1)_3(L_3)_3) + \frac{2}{3}E(FS\ M(L_2)_3(L_3)_3)) \quad (16)$$

$$E(DTS\ M(L_1)_2(L_2)_2(L_3)_2) = \frac{1}{3}(\frac{1}{3}E(HO\ M(L_1)_6) + \frac{1}{3}E(HO\ M(L_2)_6) + \frac{1}{3}E(HO\ M(L_3)_6) + \frac{2}{3}E(MS\ M(L_1)_3(L_2)_3) + \frac{2}{3}E(MS\ M(L_1)_3(L_3)_3) + \frac{2}{3}E(MS\ M(L_2)_3(L_3)_3)) \quad (17)$$

$$E(EA\ M(L_1)_2(L_2)_2(L_3)_2) = \frac{1}{3}(\frac{1}{3}E(HO\ M(L_1)_6) + \frac{1}{3}E(HO\ M(L_2)_6) + \frac{1}{3}E(HO\ M(L_3)_6) + \frac{2}{3}E(FS\ M(L_1)_3(L_2)_3) + \frac{2}{3}E(MS\ M(L_1)_3(L_3)_3) + \frac{2}{3}E(MS\ M(L_2)_3(L_3)_3)) \quad (18)$$

where the third ligand in the EA complex is the one that is *trans* to itself so energies involving that ligand are derived from the binary MS complexes, whereas the remaining components are derived from FS complexes. Analogous expressions were also obtained for the CS/TS energetics (Supporting Information Text S3). Overall, both interpolation schemes significantly improve the estimation of both spin splitting and HOMO level energies for both sets of ternary complexes in comparison to HO-only interpolation (Figures 5 and 6 and Supporting Information Figures S3 and S4 and Tables S14–S16). Differences in CA and TA complex properties are much better predicted by both interpolation schemes, with the CS/TS scheme performing best for the combination that includes $CH_3CN$ (Figures 5 and 6). While this may be expected as a generalization of observations on the binary heteroleptics, it is noteworthy that trends in the fully equivalent stoichiometries (i.e., all three EA isomers, DCS, or DTS complexes) are reasonably well predicted by the improved schemes, whereas they were indistinguishable with simple linear interpolation because they differ solely by *cis* versus *trans* positioning effects (Figures 5 and 6). Overall, errors for the interpolation scheme using FS/MS ligands are sufficiently low to warrant its use in chemical space exploration (Supporting Information Table S16). We have thus demonstrated how over a set of three ligands, nine explicit calculations can be used to interpolate the properties of 18 binary and 29 ternary complexes to around 2 kcal/mol accuracy in spin-splitting energies or 0.1–0.2 eV accuracy in orbital energy levels.

**3c. Interpolation of Chemical Space from Experimentally Characterized Complexes.**

Given the strength of the interpolative trends we observed, we chose Fe(II) complexes from the CSD to explore the potential of interpolation from previously synthesized complexes. For these 1202 complexes, the majority (661) are homoleptics, followed closely by binary complexes (407) and ternary complexes (129), with only six having more ligand types (Figure 4 and Supporting Information Table S4).We restrict our analysis of chemical space interpolation to monodentate ligands from complexes of each symmetry type that only contain monodentate ligands to simplify the interpolation process because higher-denticity ligands impose geometric constraints that make them incompatible with certain symmetry types. Nevertheless, we note that there are a significant number of monodentate ligands that only appear in combination with higher-denticity ligands (Supporting Information Tables S4–S5). From the set of monodentate-only complexes, we identified 40 unique ligands already present in homoleptic complexes along with 48 additional unique ligands present in monodentate-only binary and ternary complexes for which we could assign a charge following the scheme introduced in Ref. [60]. Of the set of 88 ligands, only 56 are assigned a neutral charge, closed-shell electronic structure, and deemed sterically feasible for homoleptic calculations (i.e., excluding only the neutral $OP(Ph)_3$ ligand).

We then computed the adiabatic spin-splitting energies and singlet HOMO levels of all 56 homoleptic Fe(II) complexes (see Computational Details). This set is strongly biased toward high-spin structures, with only a few ligands giving rise to low-spin ground states (Figure 9 and Supporting Information Figure S5 and Table S17). The majority of spin-splitting energies are in the range of -30 to 0 kcal/mol, whereas a minority of complexes are low-spin (Figure 9). Only one complex $Fe(II)(CO)_6$, which was present in our study in Sec. 3b, has a HOMO level deeper than -16 eV, and no complexes have intermediate HOMO levels (ca. -14 eV) while also sampling near degenerate spin states (Figure 9).

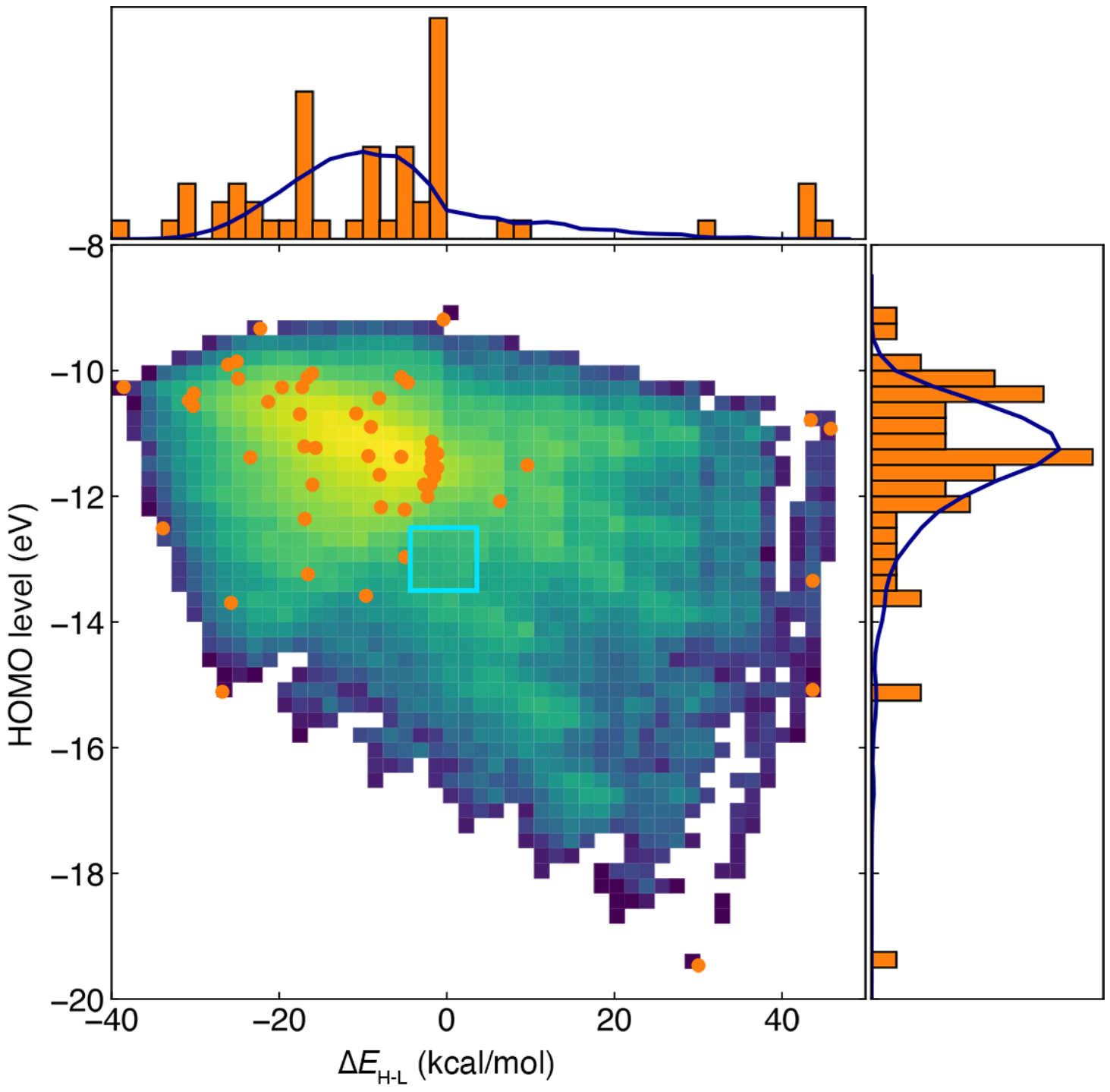

**Figure 9.** The $\Delta E_{\text{H-L}}$ (in kcal/mol) vs. singlet HOMO level (in eV) for 56 homoleptic complexes (orange circles) as well as HO-only interpolation of all possible binary and ternary complexes colored by frequency from purple (low) to yellow (high). 1D histograms of each property are shown at top and right with bin widths of 2 kcal/mol and 0.25 eV, respectively, along with a kernel density estimate of the interpolated space shown as a dark blue line. A targeted zone of -4 to 4 kcal/mol for $\Delta E_{\text{H-L}}$ and -14.0 to -13.0 eV for the HOMO is annotated as a light blue square.

We next interpolated estimates for the 816,200 binary and ternary complexes from the 56 homoleptic complexes using the HO-only averaging scheme (Figure 9). By definition, this interpolation scheme provides rough estimates of which ligand combinations will enrich $\Delta E_{\text{H-L}}$/HOMO energetic pairings not observed in the homoleptic set (Figure 9). Nevertheless, this approach provides only a coarse estimate of the energetics in comparison to interpolation schemes that use knowledge of binary complex energetics. However, even for the most data-efficient approach of FS/MS-derived interpolation, for a pool of 56 ligands, the most data-efficient approach would require 3,080 explicit FS/MS calculations to achieve high fidelity predictions for 816,200 complexes (see Table 1). Thus, we identified a way to use the HO-only interpolation to reduce the number of unique complexes required to achieve high fidelity within a target region.

First, we select a targeted region of $\Delta E_{H-L}$ values in the range of -4.0 to 4.0 kcal/mol and HOMO levels in the range of -14 to -13 eV. This region was selected because no homoleptic complexes were present in this range, but interpolation of the space predicted that heteroleptic complexes would be found there. These complexes could be of interest in chemical discovery applications that target spin-crossover candidates (i.e., with near-degenerate spin states) with good oxidative stability (i.e., deep HOMO levels). Next, we identify 12 common ligands from the parent homoleptic complexes that are predicted to give rise most frequently to binary and ternary complexes in the targeted zone on the basis of the simple HO-only model (Figure 10 and Supporting Information Table S18 and Figure S6). These ligands consist of common ligands from our original set (i.e., CO, $H_2O$, $CH_3CN$, and $NH_3$) but also introduce new chemistry (e.g, S-coordinating dimethylthioformamide, DMTF, and 2-chloropyrazine, ClPyz, Supporting Information Table S18). From this set, only 132 additional FS/MS complexes need to be studied beyond the 12 homoleptic complexes we already computed, to infer 6,776 additional HOMO level or $\Delta E_{H-L}$ properties at higher fidelity than HO-only averaging (see Table 1). We carried out geometry optimizations of these 132 complexes and used their properties to evaluate the revised interpolated HOMO level and $\Delta E_{H-L}$ values over this subset. Evaluating this subset also provides a validation of the accuracy of the homoleptic averaging over a larger set of ligands. Over this set, we observe that errors are comparable to the earlier tests (i.e., MAE of 4 kcal/mol), and homoleptic averaging is generally a good predictor of $\Delta E_{H-L}$ (Supporting Information Figure S7 and Table S19). For HOMO level predictions, there are more points that differ from the interpolated values, leading to higher MAEs (0.8 eV) than observed over our representative test ligands (Supporting Information Figure S7 and Table S19).

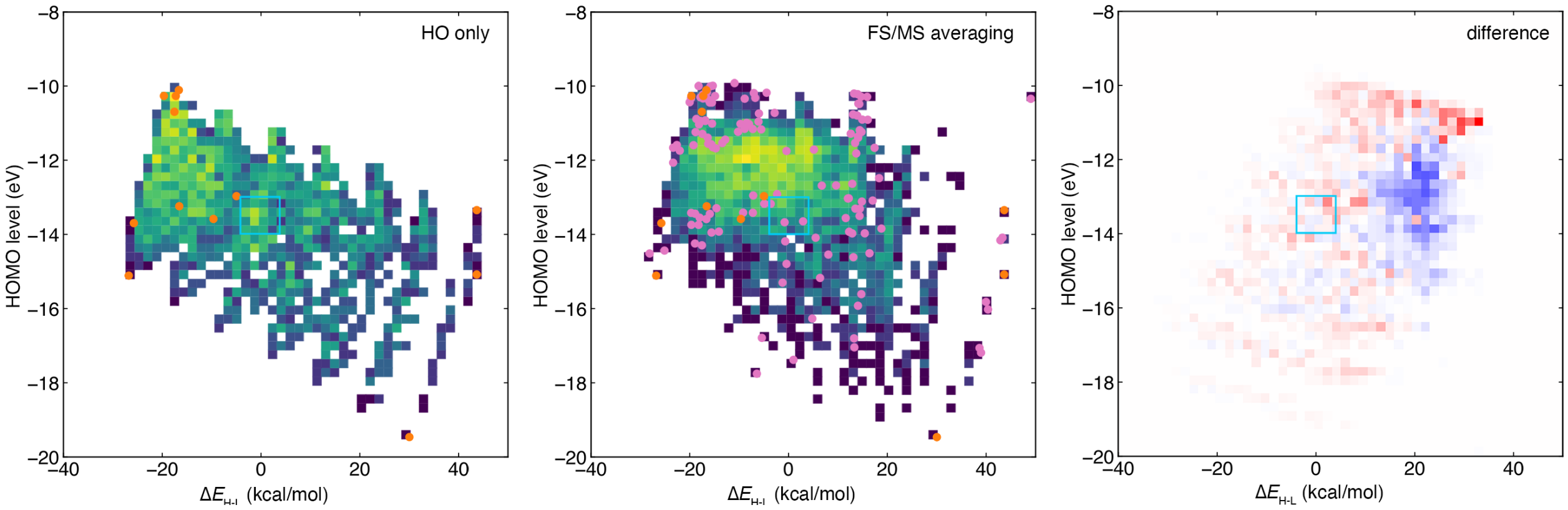

**Figure 10.** (Left and middle) The $\Delta E_{\text{H-L}}$ (in kcal/mol) vs. singlet HOMO level (in eV) for 12 homoleptic complexes (orange circles) as well as HO-only (left) or FS/MS-based (middle) interpolation of all possible binary and ternary complexes colored by frequency from purple (low) to yellow (high). The FS/MS complex energies are shown as pink circles. (Right) The difference (i.e., HO-only minus FS/MS-based interpolation) of the two 2D histograms plotted from negative (blue, -81) to positive (red, 81). A targeted zone of -4 to 4 kcal/mol for $\Delta E_{\text{H-L}}$ and -14.0 to -13.0 eV for the HOMO is annotated as a light blue square.

Nevertheless, the calculations on FS and MS complexes also highlight the limits of homoleptic averaging for seeking targeted properties. Of the 132 FS or MS calculations carried out, three are found to be in our targeted zone, MS Fe(II)(MeCN)$_3$(NH$_3$)$_3$ and both FS and MS Fe(II)(CH$_3$CN)$_3$(MeOH)$_3$ (Supporting Information Table S20). None of these three complexes were predicted to be in the zone from homoleptic averaging alone, with the MS Fe(II)(MeCN)$_3$(NH$_3$)$_3$ $\Delta E_{\text{H-L}}$ underestimated (predicted: -7.3 kcal/mol vs. actual -3.6 kcal/mol) while the FS/MS Fe(II)(CH$_3$CN)$_3$(MeOH)$_3$ $\Delta E_{\text{H-L}}$ was overestimated (predicted: 9.0 kcal/mol vs 2.9 and -0.8 kcal/mol, respectively, Supporting Information Table S20). For these same complexes, the homoleptic averaging predicted HOMO levels very well, within around 0.1 eV (i.e., lower error than for $\Delta E_{\text{H-L}}$, Supporting Information Table S20).

Returning to the properties predicted from FS/MS-interpolation, we identified a total of three additional binary complexes predicted to be in the targeted zone from FS/MS-interpolation but not in the targeted zone according to HO-only interpolation and we computed their properties (Figure 10). These complexes are 5+1 Fe(II)(MeCN)$_5$(CO) and CS/TS Fe(II)(ClPyz)$_4$(CO)$_2$ (Figure 11 and Supporting Information Table S20). Indeed, all three of these complexes have HOMO levels in the targeted zone, while the $\Delta E_{\text{H-L}}$ values are close to (4.9–5.7 kcal/mol for CS/TS Fe(II)(ClPyz)$_4$(CO)$_2$) or in the targeted zone (1.5 kcal/mol for 5+1 Fe(II)(MeCN)$_5$(CO), Supporting Information Table S20). For CS/TS Fe(II)(ClPyz)$_4$(CO)$_2$, homoleptic averaging had underestimated both the $\Delta E_{\text{H-L}}$ and HOMO level (i.e., out of the zone at -15.32 eV) and could not distinguish between CS and TS isomers (Supporting Information Table S20). Although FS/MS interpolation slightly underestimated the $\Delta E_{\text{H-L}}$ values for CS/TS Fe(II)(ClPyz)$_4$(CO)$_2$, the errors are within what we had observed over other sets. In addition to shifting which compounds are predicted to fall within the target zone, the difference between the HO-only to FS/MS interpolation schemes causes some regions of property space to be predicted to be enriched while others are predicted to be depleted (Figure 10).

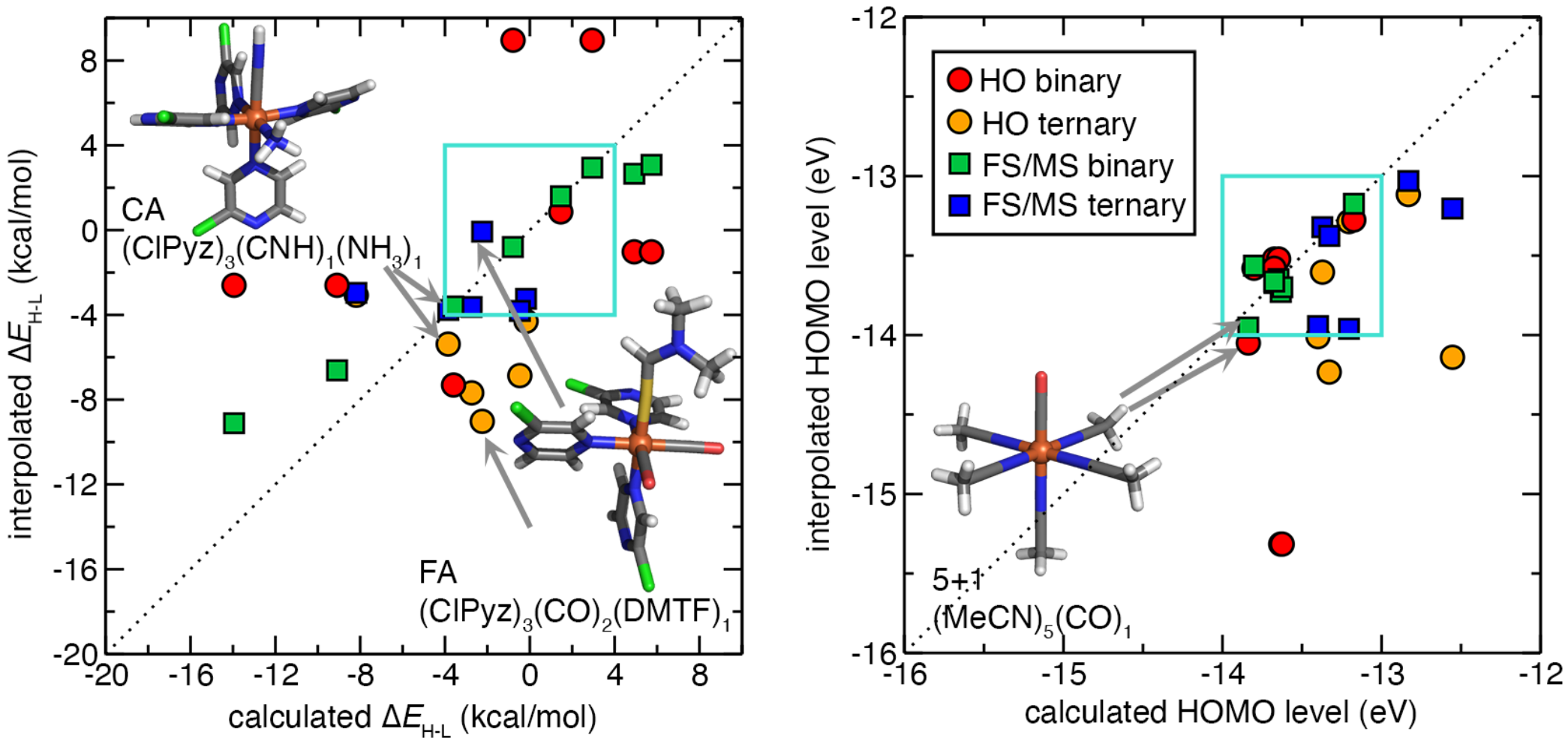

**Figure 11.** Properties of 8 binary (red or green) and 6 ternary (orange or blue) complexes in validation set for HO-interpolation (circles) and FS/MS-augmented interpolation (squares) for $\Delta E_{\text{H-L}}$ (in kcal/mol, left) and HOMO level (in eV, right). The targeted zone for each quantity is shown as a turquoise square, and three representative complexes are shown in inset with their symmetry class and the associated points are indicated with gray arrows. Structures are colored as follows: brown for Fe, gray for C, blue for N, white for H, green for Cl, red for O, and yellow for S. A dotted parity line is also shown.

As a control, we also identified two complexes predicted to be in zone by the HO-only interpolation but out of zone by FS/MS-interpolation, CS/TS $Fe(II)(MeOH)_4(CH_3CN)_2$ (Supporting Information Table S20). The homoleptic averaging predicts the $\Delta E_{H\text{-}L}$ value for both complexes to be -2.6 kcal/mol, while calculated $\Delta E_{H\text{-}L}$ values are significantly lower (-9.1 and -13.9 kcal/mol). The FS/MS interpolation captures well the relative CS/TS energetics in both this case and the in-zone example and only slightly overestimates the $\Delta E_{H\text{-}L}$ value (Supporting Information Table S20). Overall, FS/MS-interpolation MAEs for these binary complexes is low (2.5 kcal/mol for $\Delta E_{H\text{-}L}$ and 0.1 eV for HOMO levels) and less than half that observed from HO-only averaging (Supporting Information Table S20).

As a more stringent test of our FS/MS interpolation scheme, we also selected six representative ternary complexes predicted to be in the targeted zone from FS/MS interpolation but out of the targeted zone when estimated with homoleptic interpolation (Figure 11 and Supporting Information Table S21). All complexes were generally predicted to be more high-spin favoring by homoleptic averaging than predicted from FS/MS interpolation, and several were also predicted to have deeper HOMO levels than the targeted region (Supporting Information Table S21). Over this set, explicit DFT calculations show that five of the complexes have $\Delta E_{H\text{-}L}$ values in the targeted zone and four have HOMO levels in the targeted zone (Supporting Information Table S21). These include FA $Fe(II)(ClPyz)_3(CO)_2(DMTF)$, which was predicted to be strongly HS by homoleptic averaging ($\Delta E_{H\text{-}L}$ = -9.0 kcal/mol) but was much closer to the FS/MS-interpolated value (calculated $\Delta E_{H\text{-}L}$ = -2.2 kcal/mol vs FS/MS $\Delta E_{H\text{-}L}$ = -0.1 kcal/mol, Supporting Information Table S21). For the worst-performing example, FA $Fe(II)(DMTF)_3(CO)_2(H_2O)$, FS/MS interpolation overestimates $\Delta E_{H\text{-}L}$ by 5 kcal/mol (-3.0 kcal/mol vs -8.2 kcal/mol) and predicts a deeper HOMO level (-13.20 eV vs. -12.55 eV, Supporting Information Table S21). This could be due to weak coordination of the metal by water which leads to a more stabilized high-spin state. Performance of FS/MS interpolation on the remaining CA and TA complexes ranges from good (ca. 2–3 kcal/mol errors) to exceptional in the case of CA Fe(II)(ClPyz)$_4$(CNH)(NH$_3$) where errors on $\Delta E_{\text{H-L}}$ are below 0.1 kcal/mol and HOMO level are around 0.05 eV (Supporting Information Table S21). Overall errors are low from FS/MS interpolation at around 2.5 kcal/mol for $\Delta E_{\text{H-L}}$ and 0.4 eV for the HOMO level, roughly half their values from homoleptic averaging. These results demonstrate that coarse interpolation of large spaces with homoleptic complexes can be followed up by improved interpolation using selected FS/MS compounds to identify the most promising binary or ternary complexes for explicit calculation. Overall, the demonstrated approach represents a data-efficient strategy to infer properties across large compound spaces with systematically improvable fidelity.

## 4. Conclusions

A large-scale analysis of the mononuclear octahedral complexes deposited in the Cambridge Structural Database revealed a propensity towards specific, higher-symmetry classes. In addition, few complexes contained more than three unique ligand types. To assess the relative diversity of these complexes compared to the enumerated chemical space, we obtained expressions for the theoretical number of complexes of the five binary and eight ternary symmetry classes for octahedral complexes. We showed that even for a relatively small number of neutral, monodentate ligands present in Fe(II) complexes, the total theoretical space of 816,200 binary and ternary complexes far exceeded those that had been characterized in the CSD.

An aim of identifying which uncharacterized compounds are most likely to be valuable or informative motivated our evaluation of interpolative schemes to determine the extent to which heteroleptic complex properties could be inferred from parent homoleptic complexes. Over representative test cases, we observed that a linear weighted averaging of homoleptic properties could reasonably (to ca. 4 kcal/mol for $\Delta E_{\text{H-L}}$ and 0.24 eV for the HOMO level) predict properties of binary and ternary heteroleptic complexes. We demonstrated a refinement of the approach to be able to distinguish isomers (e.g., CS vs. TS or CA vs. TA) by using expressions that also incorporated either CS/TS or FS/MS binary complexes at a slightly higher computational cost but with errors that were half as large (ca. 2 kcal/mol for $\Delta E_{\text{H-L}}$ and 0.15 eV for the HOMO level). The most data-efficient approach required four FS/MS and three homoleptic energies (i.e., 7 total) to infer 18 binary and 29 ternary complex properties.

Finally, we demonstrated a two-stage discovery approach to leverage and validate our interpolative schemes. We first used 56 homoleptic Fe(II) complexes composed of neutral, closed-shell monodentate ligands to infer the properties of 816,200 binary or ternary complexes of these ligands using HO-only averaging. We then defined a targeted zone of HOMO level and $\Delta E_{\text{H-L}}$ that contained none of the homoleptic complexes. To avoid explicit calculations for all (ca. 3,000) of FS/MS complexes needed to achieve high fidelity over the full range of ligands, we then refined our analysis to the top 12 most frequently occurring ligands predicted to be in the targeted zone. From this set, we studied 66 each of FS and MS complexes to refine our interpolation of 6,776 complexes. This approach helped us to identify 3 FS/MS complexes in the targeted zone that had not been predicted by homoleptic averaging alone. It also had a higher validation rate for binary and ternary complexes than homoleptic averaging, with all FS/MS-interpolation predicted complexes residing in the targeted zone or just outside it. Overall, errors for $\Delta E_{\text{H-L}}$ of around 5 kcal/mol with homoleptic averaging and 2 kcal/mol with FS/MS-interpolation are also comparable to prior machine learning (i.e., artificial neural network) model predictions on similar data sets[51]. Thus, this approach represents a promising multi-stage strategy for efficient chemical space exploration at low cost: an initial coarse interpolation from homoleptic complexes can be systematically refined by incorporating *cis* and *trans* isomer effects over a smaller subspace of ligands. While demonstrated here for magnetic and orbital energy properties, this approach is expected to have similar applicability in predicting other properties where ligands can be expected to behave in an approximately additive manner, such as in redox potentials or catalysis. This observed additivity could also be integrated into machine learning model property predictions or used synergistically to augment datasets for machine learning.

ASSOCIATED CONTENT

**Supporting Information**. Details of mononuclear octahedral complex curation from the CSD; Details of complex symmetry class assignment; Statistics of CSD complexes by number of unique ligands; Statistics of CSD complexes by symmetry classes; Denticity of ligands in UT&C CSD complexes by symmetry classes; Denticity of ligands in Fe(II) CSD complexes by symmetry classes; Number of unique monodentate ligands in Fe(II) complexes; Sampled vs theoretical number of complexes with monodentate ligands; Homoleptic complex properties, $\Delta E_{\mathrm{H\text{-}L}}$ and HOMO level; Interpolated vs calc'd $\Delta E_{\mathrm{H\text{-}L}}$ for transition metal complexes with pairs of $NH_3$/$H_2O$/CO; Interpolated and actual $\Delta E_{\mathrm{H\text{-}L}}$ for pairs of $H_2O$/CO and $NH_3$ or $CH_3CN$; Interpolated vs calc'd HOMO for transition metal complexes with pairs of $NH_3$/$H_2O$/ CO; MAEs for $\Delta E_{\mathrm{H\text{-}L}}$ binary complex estimates from interpolation schemes; Interpolated and actual HOMO for pairs of $H_2O$/CO and $NH_3$ or $CH_3CN$; MAEs for HOMO binary complex estimates from interpolation schemes; Interpolated vs calc'd $\Delta E_{\mathrm{H\text{-}L}}$ for transition metal complexes with three ligands: $NH_3$/$H_2O$/CO; Interpolated and actual $\Delta E_{\mathrm{H\text{-}L}}$ for ternary $H_2O$/CO/$CH_3CN$ complexes; Interpolated and actual $\Delta E_{\mathrm{H\text{-}L}}$ for ternary $H_2O$/CO/$NH_3$ complexes; Interpolated vs calc'd HOMO for transition metal complexes with three ligands: $NH_3$/$H_2O$/CO; Interpolated and actual HOMO for ternary $H_2O$/CO/$CH_3CN$ complexes; Interpolated and actual HOMO for ternary $H_2O$/CO/$NH_3$ complexes; CS/TS-derived interpolation expressions for ternary complexes; MAEs of HOMO and $\Delta E_{\mathrm{H\text{-}L}}$ for ternary complexes; Properties of 52 homoleptic Fe(II) complexes; Scatter of HOMO and $DE_{\mathrm{H\text{-}L}}$ for 52 homoleptic Fe(II) complexes; 12 most common ligands in target zone for Fe(II) complexes; Homoleptic-only interpolated target zone for Fe(II) complexes; Parity plot for predicted vs calculated 132 FS/MS Fe(II) complexes; MAEs for FS and MS complexes from HO-only averaging; Binary complexes in validation set for interpolation; Ternary complexes in validation set for interpolation. (PDF)

Pickle file of CSD structures; xyz files of all structures studied; csv file of energies and properties of complexes studied. (ZIP)

AUTHOR INFORMATION

**Corresponding Author**

*email:hjkulik@mit.edu

**Notes**

The authors declare no competing financial interest.

ACKNOWLEDGMENT

This work was primarily supported by the Office of Naval Research under grant number N00014-20-1-2150. A.N., N.A., and D.B.K.C. were partially supported by National Science Foundation Graduate Research Fellowships under Grants #1122374 (to A.N.) and #1745302 (to N.A. and D.B.K.C). C.D. was partially supported by a seed fellowship from the Molecular Sciences Software Institute under NSF grant OAC-1547580. The authors acknowledge Adam H. Steeves for a critical reading of the manuscript.

REFERENCES

1. Y. N. Shu and B. G. Levine, J Chem Phys **142** (10), 104104 (2015).
2. R. Gomez-Bombarelli, J. Aguilera-Iparraguirre, T. D. Hirzel, D. Duvenaud, D. Maclaurin, M. A. Blood-Forsythe, H. S. Chae, M. Einzinger, D. G. Ha, T. Wu, G. Markopoulos, S. Jeon, H. Kang, H. Miyazaki, M. Numata, S. Kim, W. L. Huang, S. I. Hong, M. Baldo, R. P. Adams and A. Aspuru-Guzik, Nat Mater **15** (10), 1120-+ (2016).
3. I. Y. Kanal, S. G. Owens, J. S. Bechtel and G. R. Hutchison, J. Phys. Chem. Lett. **4** (10), 1613-1623 (2013).
4. K. D. Vogiatzis, M. V. Polynski, J. K. Kirkland, J. Townsend, A. Hashemi, C. Liu and E. A. Pidko, Chemical Reviews **119**, 2453-2523 (2018).
5. M. Foscato and V. R. Jensen, ACS Catal. **10** (3), 2354-2377 (2020).
6. S. Curtarolo, G. L. Hart, M. B. Nardelli, N. Mingo, S. Sanvito and O. Levy, Nat Mater **12** (3), 191-201 (2013).
7. S. P. Ong, W. D. Richards, A. Jain, G. Hautier, M. Kocher, S. Cholia, D. Gunter, V. L. Chevrier, K. A. Persson and G. Ceder, Computational Materials Science **68**, 314-319 (2013).
8. J. K. Nørskov and T. Bligaard, Angewandte Chemie International Edition **52** (3), 776-777 (2013).
9. J. P. Janet, C. Duan, A. Nandy, F. Liu and H. J. Kulik, Accounts of Chemical Research **54** (3), 532-545 (2021).

10. M. Ceriotti, C. Clementi and O. A. von Lilienfeld, Chemical Reviews **121** (16), 9719-9721 (2021).
11. J. A. Keith, V. Vassilev-Galindo, B. Q. Cheng, S. Chmiela, M. Gastegger, K. R. Mueller and A. Tkatchenko, Chemical Reviews **121** (16), 9816-9872 (2021).
12. M. Ceriotti, J Chem Phys **150** (15), 150901 (2019).
13. A. Grisafi, A. Fabrizio, B. Meyer, D. M. Wilkins, C. Corminboeuf and M. Ceriotti, ACS Central Science **5**, 57-64 (2018).
14. K. Jorner, A. Tomberg, C. Bauer, C. Skold and P. O. Norrby, Nat Rev Chem **5** (4), 240-255 (2021).
15. M. F. Kasim and S. M. Vinko, Phys Rev Lett **127** (12), 126403 (2021).
16. J. Westermayr, M. Gastegger, K. T. Schutt and R. J. Maurer, J Chem Phys **154** (23), 230903 (2021).
17. J. P. Coe, J Chem Theory Comput **14** (11), 5739-5749 (2018).
18. A. S. Rosen, S. M. Iyer, D. Ray, Z. P. Yao, A. Aspuru-Guzik, L. Gagliardi, J. M. Notestein and R. Q. Snurr, Matter-Us **4** (5), 1578-1597 (2021).
19. T. Dimitrov, C. Kreisbeck, J. S. Becker, A. Aspuru-Guzik and S. K. Saikin, ACS Appl. Mater. Interfaces **11** (28), 24825-24836 (2019).
20. A. Glielmo, B. E. Husic, A. Rodriguez, C. Clementi, F. Noe and A. Laio, Chemical Reviews **121** (16), 9722-9758 (2021).
21. L. Chanussot, A. Das, S. Goyal, T. Lavril, M. Shuaibi, M. Riviere, K. Tran, J. Heras-Domingo, C. Ho, W. H. Hu, A. Palizhati, A. Sriram, B. Wood, J. Yoon, D. Parikh, C. L. Zitnick and Z. Ulissi, Acs Catalysis **11** (10), 6059-6072 (2021).
22. M. Christensen, L. P. E. Yunker, F. Adedeji, F. Hase, L. M. Roch, T. Gensch, G. D. Gomes, T. Zepel, M. S. Sigman, A. Aspuru-Guzik and J. E. Hein, Commun Chem **4** (1), 112 (2021).
23. T. Gensch, G. D. Gomes, P. Friederich, E. Peters, T. Gaudin, R. Pollice, K. Jorner, A. Nigam, M. Lindner-D'Addario, M. S. Sigman and A. Aspuru-Guzik, J Am Chem Soc **144** (3), 1205-1217 (2022).
24. A. M. Virshup, J. Contreras-García, P. Wipf, W. Yang and D. N. Beratan, J. Am. Chem. Soc. **135** (19), 7296-7303 (2013).
25. L. Ruddigkeit, R. van Deursen, L. C. Blum and J.-L. Reymond, J. Chem. Inf. Model. **52** (11), 2864-2875 (2012).
26. R. W. Sterner and J. J. Elser, *Ecological stoichiometry: the biology of elements from molecules to the biosphere*. (Princeton University Press, 2002).
27. H. J. M. Bowen, *Environmental chemistry of the elements*. (Academic Press., 1979).
28. T. Fink, H. Bruggesser and J.-L. Reymond, Angewandte Chemie International Edition **44** (10), 1504-1508 (2005).
29. M. J. Wester, S. N. Pollock, E. A. Coutsias, T. K. Allu, S. Muresan and T. I. Oprea, J. Chem. Inf. Model. **48** (7), 1311-1324 (2008).
30. P. G. Polishchuk, T. I. Madzhidov and A. Varnek, Journal of computer-aided molecular design **27** (8), 675-679 (2013).
31. R. S. Bohacek, C. McMartin and W. C. Guida, Medicinal research reviews **16** (1), 3-50 (1996).
32. N. Fey, S. E. Harris, J. N. Harvey and A. G. Orpen, J. Chem. Inf. Model. **46** (2), 912-929 (2006).

33. N. Fey, A. C. Tsipis, S. E. Harris, J. N. Harvey, A. G. Orpen and R. A. Mansson, Chem. - Eur. J. **12** (1), 291-302 (2006).
34. D. J. Durand and N. Fey, Chem. Rev. **119** (11), 6561-6594 (2019).
35. M. Foscato, G. Occhipinti, V. Venkatraman, B. K. Alsberg and V. R. Jensen, J. Chem. Inf. Model. **54** (3), 767-780 (2014).
36. M. Foscato, V. Venkatraman, G. Occhipinti, B. K. Alsberg and V. R. Jensen, J. Chem. Inf. Model. **54** (7), 1919-1931 (2014).
37. M. Foscato, R. J. Deeth and V. R. Jensen, J. Chem. Inf. Model. **55** (6), 1282-1290 (2015).
38. M. Foscato, B. J. Houghton, G. Occhipinti, R. J. Deeth and V. R. Jensen, J. Chem. Inf. Model. **55** (9), 1844-1856 (2015).
39. D. Balcells and B. B. Skjelstad, J. Chem. Inf. Model. **60** (12), 6135-6146 (2020).
40. J. S. Smith, O. Isayev and A. E. Roitberg, Sci. Data **4**, 170193 (2017).
41. R. Ramakrishnan, P. O. Dral, M. Rupp and O. A. Von Lilienfeld, Sci. Data **1**, 140022 (2014).
42. C. A. Gaggioli, S. J. Stoneburner, C. J. Cramer and L. Gagliardi, ACS catalysis **9** (9), 8481-8502 (2019).
43. W. Jiang, N. J. DeYonker and A. K. Wilson, J. Chem. Theory Comput. **8** (2), 460-468 (2012).
44. F. Liu, C. Duan and H. J. Kulik, J. Phys. Chem. Lett. **11**, 8067-8076 (2020).
45. C. Duan, D. B. K. Chu, A. Nandy and H. J. Kulik, Chem. Sci. **13**, 4962-4971 (2022).
46. M. Feldt, Q. M. Phung, K. Pierloot, R. A. Mata and J. N. Harvey, J. Chem. Theory Comput. **15** (2), 922-937 (2019).
47. Q. M. Phung, M. Feldt, J. N. Harvey and K. Pierloot, J. Chem. Theory Comput. **14** (5), 2446-2455 (2018).
48. S. Gugler, J. P. Janet and H. J. Kulik, Mol. Syst. Des. Eng. **5**, 139-152 (2020).
49. A. Gaulton, A. Hersey, M. Nowotka, A. P. Bento, J. Chambers, D. Mendez, P. Mutowo, F. Atkinson, L. J. Bellis, E. Cibrián-Uhalte, M. Davies, N. Dedman, A. Karlsson, M. P. Magariños, J. P. Overington, G. Papadatos, I. Smit and A. R. Leach, Nucleic Acids Res **45** (D1), D945-D954 (2017).
50. P. F. Bernath and S. McLeod, J Mol Spectrosc **207** (2), 287 (2001).
51. J. P. Janet, C. Duan, T. Yang, A. Nandy and H. J. Kulik, Chem. Sci. **10**, 7913-7922 (2019).
52. C. R. Groom, I. J. Bruno, M. P. Lightfoot and S. C. Ward, Acta Crystallographica Section B Structural Science, Crystal Engineering and Materials **72** (2), 171-179 (2016).
53. J. Glerup, O. Monsted and C. E. Schaffer, Inorg Chem **15** (6), 1399-1407 (1976).
54. R. J. Deeth, D. L. Foulis and B. J. Williams-Hubbard, Dalton Trans. (20), 3949-3955 (2003).
55. Y. Cytter, A. Nandy, A. Bajaj and H. J. Kulik, The Journal of Physical Chemistry Letters **13**, 4549-4555 (2022).
56. R. A. Friesner, E. H. Knoll and Y. Cao, J. Chem. Phys. **125** (12), 124107 (2006).
57. D. Rinaldo, L. Tian, J. N. Harvey and R. A. Friesner, J. Chem. Phys. **129** (16), 164108 (2008).
58. T. F. Hughes and R. A. Friesner, J. Chem. Theory Comput. **7** (1), 19-32 (2011).
59. R. A. Friesner and S. V. Jerome, Coord. Chem. Rev. **344**, 205-213 (2017).
60. C. Duan, A. J. Ladera, J. C.-L. Liu, M. G. Taylor, I. R. Ariyarathna and H. J. Kulik, J Chem Theory Comput (in press).

61. K. U. Lao and J. M. Herbert, The Journal of Physical Chemistry A **119** (2), 235-252 (2015).
62. M. S. Gordon, L. Slipchenko, H. Li and J. H. Jensen, Annual reports in computational chemistry **3**, 177-193 (2007).
63. F. Neese, A. Hansen and D. G. Liakos, The Journal of chemical physics **131** (6), 064103 (2009).
64. Q. Ma and H. J. Werner, Wiley Interdisciplinary Reviews: Computational Molecular Science **8** (6), e1371 (2018).
65. M. G. Taylor, T. Yang, S. Lin, A. Nandy, J. P. Janet, C. Duan and H. J. Kulik, The Journal of Physical Chemistry A **124** (16), 3286-3299 (2020).
66. S. Seritan, C. Bannwarth, B. S. Fales, E. G. Hohenstein, S. I. Kokkila-Schumacher, N. Luehr, J. W. Snyder Jr, C. Song, A. V. Titov and I. S. Ufimtsev, The Journal of chemical physics **152** (22), 224110 (2020).
67. S. Seritan, C. Bannwarth, B. S. Fales, E. G. Hohenstein, C. M. Isborn, S. I. Kokkila-Schumacher, X. Li, F. Liu, N. Luehr and J. W. Snyder Jr, Wiley Interdisciplinary Reviews: Computational Molecular Science **11** (2), e1494 (2021).
68. A. D. Becke, The Journal of Chemical Physics **98** (7), 5648-5652 (1993).
69. C. Lee, W. Yang and R. G. Parr, Physical Review B **37**, 785--789 (1988).
70. P. J. Stephens, F. J. Devlin, C. F. Chabalowski and M. J. Frisch, The Journal of Physical Chemistry **98** (45), 11623-11627 (1994).
71. P. J. Hay and W. R. Wadt, The Journal of Chemical Physics **82** (1), 270-283 (1985).
72. R. Ditchfield, W. J. Hehre and J. A. Pople, J Chem Phys **54** (2), 724 (1971).
73. V. R. Saunders and I. H. Hillier, International Journal of Quantum Chemistry **7** (4), 699-705 (1973).
74. L.-P. Wang and C. Song, The Journal of Chemical Physics **144** (21), 214108 (2016).
75. E. I. Ioannidis, T. Z. H. Gani and H. J. Kulik, J. Comput. Chem. **37**, 2106-2117 (2016).
76. KulikGroup.
77. N. M. O'Boyle, M. Banck, C. A. James, C. Morley, T. Vandermeersch and G. R. Hutchison, J Cheminformatics **3**, 33 (2011).
78. N. M. O'Boyle, C. Morley and G. R. Hutchison, Chemistry Central Journal **2**, 5 (2008).
79. G. Pólya and R. C. Read, (1987).
80. I. Baraldi, C. Fiori and D. Vanossi, 8.
81. B. A. Kennedy, D. A. McQuarrie and C. H. Brubaker, Inorg Chem **3** (2), 265-268 (1964).
82. A. Nandy, D. B. K. Chu, D. R. Harper, C. Duan, N. Arunachalam, Y. Cytter and H. J. Kulik, Phys. Chem. Chem. Phys. **22**, 19326-19341 (2020).

**Supporting Information for**

***Ligand additivity relationships enable efficient exploration of transition metal chemical space***

Naveen Arunachalam[1,#], Stefan Gugler[1,#], Michael G. Taylor[1,#], Chenru Duan[1,2], Aditya Nandy[1,2], Jon Paul Janet[1], Ralf Meyer[1], Jonas Oldenstaedt[1], Daniel B. K. Chu[1], and Heather J. Kulik[1,*]
[1]Department of Chemical Engineering, Massachusetts Institute of Technology, Cambridge, MA 02139, USA
[2]Department of Chemistry, Massachusetts Institute of Technology, Cambridge, MA 02139, USA
*email: hjkulik@mit.edu

**Contents**

**Text S1.** Description of mononuclear complex curation.
The Conquest interface was used to query for structures containing a transition metal atom (here, groups 3–12, periods 4–6, excluding La) that forms at least 4 bonds with either p-block (here, the first five rows of groups 13−17), noble gas, or group 1 or 2 elements. The only constraint imposed on the search was that 3D coordinates were determined. This initial query produced 462,012 Refcodes for unique structures. The CSD Python API was then used to iterate through every "component" molecule in each x-ray crystallographic structure to identify all non-polymeric molecules with distinct molecular weight (within the same crystal structure) and only one transition metal atom. For each of the resulting 240,117 mononuclear transition metal complexes the CSD mol2 files were saved after adding any missing hydrogen atoms, preserving both the connectivity and bond orders present in the CSD. From this diverse set, 6-coordinate metal centers without edge-bound or sandwich-type ligands were evaluated for deviation of the angles (as defined in Ref. 1) between the metal-connecting atoms from the VSEPR ideal angles for octahedral, pentagonal pyramidal, or trigonal prismatic structures. Structures with lowest angular deviation from ideal octahedral molecular structure over other geometries were identified as octahedral.

**Text S2.** Description of complex symmetry identification.

Within each complex, a 3-hydrogen difference criterion was introduced to heuristically capture ligands that only differed in protonation state. If none of the equivalence conditions were met the ligands were identified as chemically identical. Ligand symmetry was identified by binning the octahedral structures by: (1) Ligand denticities (e.g., 5-denticity + 1-denticity) (2) maximum number of connecting atoms from a single ligand in the equatorial plane (e.g., 3-denticity + 3-denticity has at most 2 connecting atoms from the same ligand in the equatorial plane -> *fac* binding) (3) total number of unique ligands (e.g., for a monodentate-only complex with 5 unique ligands, two must be duplicates) and (4) for lower denticities, counts of pairs of *trans* ligands that are identical (e.g., for a monodentate-only complex with 3 pairs of identical ligands, if only 1 pair of *trans* ligands are identical -> other unique pairs must be *cis*). A nomenclature for ligand symmetry was created to describe the observed symmetries following these rules (1) Denticities are denoted by numbers as the base of the ligand symmetry representation with higher denticities listed first. (2) If ligands are identical they are surrounded by "||" markers, with ligands with the highest number of identical copies of the same denticity listed first. (3) Tridentate ligands are identified as either *fac*- or *mer*- by f/m labels, and corresponding sets of three identical monodentate ligands are given equivalent f/m labels. (4) Pairs of identical monodentate ligands are identified as either *cis*- or *trans*- with c/t labels, pairs of non-identical monodentate ligands are given c/t labels only when there is one pair of non-identical monodentate ligands, and the ligand symmetry of identical pairs is ambiguous. Examples of ligand symmetry using this nomenclature are "|111111|" for a homoleptic monodentate-only complex, "4|11|t" for an equatorial tetradentate with *trans* identical monodentate ligands, and "2|11|c11" for a complex with a bidentate and 2 identical monodentate ligands in the equatorial plane with 2 chemically-distinct monodentate ligands in the axial positions.

**Table S1.** Number of complexes grouped by number of unique ligands for all (all) mononuclear octahedral transition metal complexes, all unique complexes with user-defined charges and non-disordered structures ("computation-ready"), and the subset of unique complexes with Fe(II) centers (Fe(II)). The percent of each category is also shown.

| # ligands | all | percent | computation-ready | percent | Fe(II) | percent |
|---|---|---|---|---|---|---|
| 1 | 19324 | 22.6% | 4830 | 28.3% | 661 | 55.0% |
| 2 | 38780 | 45.3% | 8191 | 47.9% | 407 | 33.8% |
| 3 | 22655 | 26.5% | 3701 | 21.7% | 129 | 10.7% |
| 4 | 4530 | 5.3% | 350 | 2.1% | 5 | 0.4% |
| 5 | 283 | 0.3% | 13 | 0.1% | 0 | 0.0% |
| 6 | 3 | 0.0% | 0 | 0.0% | 0 | 0.0% |
| total | 85575 | | 17085 | | 1202 | |

**Table S2.** Number of complexes grouped by symmetry class for all mononuclear octahedral transition metal complexes, all complexes with user-defined charges and non-disordered structures ("computation-ready"), and the subset of complexes with Fe(II) centers. The percent of each category is also shown.

| Symmetry class | All | percent | computation-ready | percent | Fe(II) | percent |
|---|---|---|---|---|---|---|
| HO | 19324 | 24% | 4830 | 29% | 661 | 55% |
| 5+1 | 4881 | 6% | 1216 | 7% | 128 | 11% |
| CS | 15138 | 19% | 2447 | 15% | 74 | 6% |
| TS | 14861 | 18% | 3745 | 22% | 151 | 13% |
| FS | 2184 | 3% | 426 | 3% | 36 | 3% |
| MS | 1716 | 2% | 357 | 2% | 18 | 2% |
| CA | 2571 | 3% | 793 | 5% | 26 | 2% |
| TA | 3691 | 5% | 760 | 5% | 41 | 3% |
| FA | 5828 | 7% | 772 | 5% | 21 | 2% |
| MAC | 4706 | 6% | 826 | 5% | 17 | 1% |
| MAT | 1821 | 2% | 205 | 1% | 14 | 1% |
| DCS | 802 | 1% | 113 | 1% | 2 | 0% |
| DTS | 1011 | 1% | 96 | 1% | 4 | 0% |
| EA | 2225 | 3% | 136 | 1% | 4 | 0% |
| Total | 80759 | | 16722 | | 1197 | |

**Table S3.** Number of unique ligands of each denticity grouped by symmetry class for all complexes with user-defined charges and non-disordered structures.

| | | # unique ligands by denticity | | | | | | |
|---|---|---|---|---|---|---|---|---|
| | # complexes | 1 | 2 | 3 | 4 | 5 | 6 | total |
| HO | 4830 | 150 | 528 | 1719 | 0 | 0 | 813 | 3210 |
| 5+1 | 1216 | 242 | 0 | 0 | 0 | 507 | 0 | 749 |
| TS | 3745 | 677 | 786 | 0 | 700 | 0 | 0 | 2163 |
| CS | 2447 | 109 | 1342 | 0 | 295 | 0 | 0 | 1746 |
| FS | 426 | 48 | 0 | 275 | 0 | 0 | 0 | 323 |
| MS | 357 | 35 | 0 | 318 | 0 | 0 | 0 | 353 |
| CA | 793 | 302 | 119 | 0 | 206 | 0 | 0 | 627 |
| DCS | 113 | 27 | 105 | 0 | 0 | 0 | 0 | 132 |
| TA | 760 | 321 | 66 | 165 | 0 | 0 | 0 | 552 |
| DTS | 96 | 111 | 0 | 0 | 0 | 0 | 0 | 111 |
| EA | 136 | 66 | 118 | 0 | 0 | 0 | 0 | 184 |
| FA | 772 | 225 | 226 | 168 | 0 | 0 | 0 | 619 |
| MAC | 826 | 195 | 266 | 252 | 0 | 0 | 0 | 713 |
| MAT | 205 | 98 | 1 | 121 | 0 | 0 | 0 | 220 |

**Table S4.** Number of unique ligands of each denticity grouped by symmetry class for all complexes with user-defined Fe(II) centers.

| | | # unique ligands by denticity | | | | | | |
|---|---|---|---|---|---|---|---|---|
| | # complexes | 1 | 2 | 3 | 4 | 5 | 6 | total |
| HO | 661 | 40 | 142 | 355 | 0 | 0 | 124 | 661 |
| 5+1 | 128 | 29 | 0 | 0 | 0 | 81 | 0 | 110 |
| TS | 151 | 53 | 71 | 0 | 49 | 0 | 0 | 173 |
| CS | 74 | 18 | 27 | 0 | 44 | 0 | 0 | 89 |
| FS | 36 | 11 | 0 | 24 | 0 | 0 | 0 | 35 |
| MS | 18 | 6 | 0 | 22 | 0 | 0 | 0 | 28 |
| CA | 26 | 18 | 6 | 0 | 17 | 0 | 0 | 41 |
| DCS | 2 | 3 | 3 | 0 | 0 | 0 | 0 | 6 |
| TA | 41 | 34 | 7 | 0 | 10 | 0 | 0 | 51 |
| DTS | 4 | 8 | 0 | 0 | 0 | 0 | 0 | 8 |
| EA | 4 | 5 | 3 | 0 | 0 | 0 | 0 | 8 |
| FA | 21 | 16 | 9 | 14 | 0 | 0 | 0 | 39 |
| MAC | 17 | 14 | 6 | 9 | 0 | 0 | 0 | 29 |
| MAT | 14 | 13 | 0 | 11 | 0 | 0 | 0 | 24 |

**Table S5.** Number of monodentate ligands of each category from Fe(II) monodentate-only structures for the case where user defined charges and trustworthy structures were obtained ("All computation-ready") as well as the more stringent test where ligand charges were assigned as neutral. The bolded categories correspond to those used to study HO Fe(II) complexes as described in the main text.

| **Category** | **All computation-ready** | **neutral ligand charge** |
|---|---|---|
| HO | 40 | 36 |
| **2, or 3 type but not HO** | **48** | **21** |
| **Total 1, 2, or 3 types** | 88 | 57 |
| only in 4-5-6-types | 3 | 1 |

**Table S6.** Number of complexes from monodentate-only configurations for each symmetry from the total user defined charges and trustworthy structures ("computation-ready") or for Fe(II) only. For Fe(II) complexes, theoretical spaces of $N$ = 56 and $N$ = 88 ligands are compared for each symmetry class along with the % of that space present in the CSD. The relevant symmetry class and notation is also provided.

| | | | | | ***N* = 88** | | ***N* = 56** | | |
|---|---|---|---|---|---|---|---|---|---|
| | **Computation-ready** | **Percent** | **Fe(II)** | **Percent** | **#** | **%** | **#** | **%** | **Symmetry class** |
| HO | 150 | 7% | 40 | 39% | 88 | 45.45% | 56 | 71.43% | l111111l |
| 5+1 | 330 | 16% | 9 | 9% | 7656 | 0.12% | 3080 | 0.29% | l11111l1 |
| TS | 944 | 46% | 26 | 25% | 7656 | 0.34% | 3080 | 0.84% | l1111ll11lt |
| CS | 92 | 5% | 5 | 5% | 7656 | 0.07% | 3080 | 0.16% | l1111ll11lc |
| FS | 48 | 2% | 5 | 5% | 3828 | 0.13% | 1540 | 0.32% | l111lfl111l |
| MS | 23 | 1% | 0 | 0% | 3828 | 0.00% | 1540 | 0.00% | l111lml111l |
| CA | 46 | 2% | 0 | 0% | 329208 | 0.00% | 83160 | 0.00% | l1111l11c |
| DCS | 0 | 0% | 0 | 0% | 109736 | 0.00% | 27720 | 0.00% | l11lcl11lcl11lc |
| TA | 183 | 9% | 3 | 3% | 329208 | 0.00% | 83160 | 0.00% | l1111l11t |
| DTS | 96 | 5% | 4 | 4% | 109736 | 0.00% | 27720 | 0.01% | l11ltl11ltl11lt |
| EA | 17 | 1% | 1 | 1% | 329208 | 0.00% | 83160 | 0.00% | l11lcl11lcl11lt |
| FA | 36 | 2% | 4 | 4% | 658416 | 0.00% | 166,320 | 0.00% | l111lfl11l1 |
| MAC | 35 | 2% | 3 | 3% | 658416 | 0.00% | 166,320 | 0.00% | l111lml11lc1 |
| MAT | 39 | 2% | 2 | 2% | 658416 | 0.00% | 166,320 | 0.00% | l111lml11lt1 |
| Total | 2039 | | 102 | | 3213056 | 0.00% | 816,256 | 0.02% | |

**Table S7.** Homoleptic Fe(II) transition metal complex properties, $\Delta E_{\text{H-L}}$ (in kcal/mol) and singlet HOMO level (in eV) for the ligands $NH_3$, $CH_3CN$, CO, and $H_2O$ obtained with B3LYP/LACVP* calculations.

| **Ligand** | **$\Delta E_{\text{H-L}}$ (kcal/mol)** | **HOMO level (eV)** |
|---|---|---|
| $CH_3CN$ | 43.67 | -13.35 |
| CO | 30.34 | -19.46 |
| $H_2O$ | -27.04 | -15.16 |
| $NH_3$ | -9.64 | -13.58 |

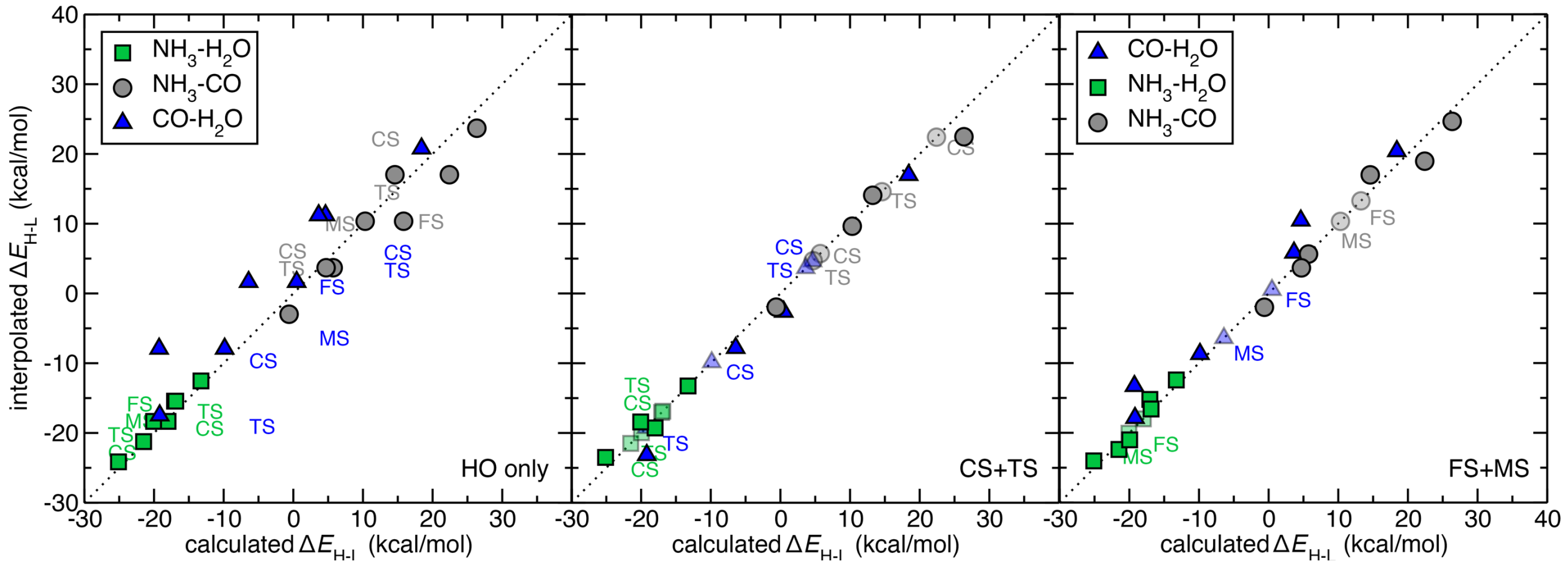

**Figure S1.** Calculated vs. linearly interpolated $\Delta E_{\text{H-L}}$ (kcal/mol) for Fe(II) complexes with pairs of any of the three ligands: $NH_3$, $H_2O$, and CO. From left to right: interpolation between homoleptic complexes (HO only), interpolation using homoleptic complexes as well as *cis* symmetric and *trans* symmetric complex energies (CS+TS), or interpolation using homoleptic complexes as well as *fac* and *mer* symmetric complex energies (FS+MS). Points are colored according to the pair of ligands they correspond to: $NH_3$-$H_2O$ (green squares), $NH_3$-CO (gray circles), and CO-$H_2O$ (blue triangles), as indicated in inset legend. Key isomers are annotated. Points provided for the fit are translucent, whereas the remaining points are opaque. In all panes, a black dotted parity line is shown.

**Table S8.** $\Delta E_{\text{H-L}}$ values for pairs of two ligand types in Fe(II) complexes (in kcal/mol) along with interpolated values using only HO complexes, HO+CS/TS complexes, or HO+FS/MS complexes.

| | | | Number of ligands | | | | | interpolated | | |
|---|---|---|---|---|---|---|---|---|---|---|
| **$L_1$** | **$L_2$** | **Symmetry** | **CO** | **$H_2O$** | **$CH_3CN$** | **$NH_3$** | **splitting** | **HO** | **CS/TS** | **FS/MS** |
| $H_2O$ | $CH_3CN$ | 5+1 | 0 | 5 | 1 | 0 | -18.4 | -15.3 | -19.9 | -16.2 |
| $H_2O$ | $CH_3CN$ | CS | 0 | 4 | 2 | 0 | -7.8 | -3.5 | -7.8 | -5.4 |
| $H_2O$ | $CH_3CN$ | TS | 0 | 4 | 2 | 0 | -12.8 | -3.5 | -12.8 | -9.6 |
| $H_2O$ | $CH_3CN$ | FS | 0 | 3 | 3 | 0 | 5.4 | 8.3 | 2.8 | 5.4 |
| $H_2O$ | $CH_3CN$ | MS | 0 | 3 | 3 | 0 | -0.9 | 8.3 | -1.2 | -0.9 |
| $CH_3CN$ | $H_2O$ | CS | 0 | 2 | 4 | 0 | 13.4 | 20.1 | 13.4 | 18.2 |
| $CH_3CN$ | $H_2O$ | TS | 0 | 2 | 4 | 0 | 10.5 | 20.1 | 10.5 | 14.0 |
| $CH_3CN$ | $H_2O$ | 5+1 | 0 | 1 | 5 | 0 | 26.3 | 31.9 | 27.1 | 30.9 |
| $CH_3CN$ | CO | 5+1 | 1 | 0 | 5 | 0 | 42.6 | 41.4 | 42.5 | 41.9 |
| $CH_3CN$ | CO | CS | 2 | 0 | 4 | 0 | 39.2 | 39.2 | 39.2 | 40.1 |
| $CH_3CN$ | CO | TS | 2 | 0 | 4 | 0 | 41.2 | 39.2 | 41.2 | 40.0 |
| CO | $CH_3CN$ | FS | 3 | 0 | 3 | 0 | 38.4 | 37.0 | 39.0 | 38.4 |
| CO | $CH_3CN$ | MS | 3 | 0 | 3 | 0 | 38.2 | 37.0 | 40.2 | 38.2 |
| CO | $CH_3CN$ | CS | 4 | 0 | 2 | 0 | 38.8 | 34.8 | 38.8 | 35.7 |
| CO | $CH_3CN$ | TS | 4 | 0 | 2 | 0 | 39.1 | 34.8 | 39.1 | 35.6 |
| CO | $CH_3CN$ | 5+1 | 5 | 0 | 1 | 0 | 35.0 | 32.6 | 34.7 | 33.0 |
| $H_2O$ | CO | 5+1 | 1 | 5 | 0 | 0 | -19.2 | -17.5 | -23.2 | -17.9 |
| $H_2O$ | CO | CS | 2 | 4 | 0 | 0 | -9.9 | -7.9 | -9.9 | -8.7 |
| $H_2O$ | CO | TS | 2 | 4 | 0 | 0 | -19.3 | -7.9 | -19.3 | -13.3 |
| $H_2O$ | CO | FS | 3 | 3 | 0 | 0 | 0.5 | 1.6 | -2.6 | 0.5 |
| $H_2O$ | CO | MS | 3 | 3 | 0 | 0 | -6.4 | 1.6 | -7.8 | -6.4 |
| CO | $H_2O$ | CS | 4 | 2 | 0 | 0 | 4.6 | 11.2 | 4.6 | 10.4 |
| CO | $H_2O$ | TS | 4 | 2 | 0 | 0 | 3.6 | 11.2 | 3.6 | 5.8 |
| CO | $H_2O$ | 5+1 | 5 | 1 | 0 | 0 | 18.4 | 20.8 | 17.0 | 20.4 |
| $NH_3$ | $H_2O$ | 5+1 | 0 | 1 | 0 | 5 | -13.3 | -12.5 | -13.3 | -12.4 |
| $NH_3$ | $H_2O$ | CS | 0 | 2 | 0 | 4 | -17.1 | -15.4 | -17.1 | -15.2 |
| $NH_3$ | $H_2O$ | TS | 0 | 2 | 0 | 4 | -16.9 | -15.4 | -16.9 | -16.6 |
| $H_2O$ | $NH_3$ | FS | 0 | 3 | 0 | 3 | -18.0 | -18.3 | -19.3 | -18.0 |
| $H_2O$ | $NH_3$ | MS | 0 | 3 | 0 | 3 | -20.1 | -18.3 | -18.4 | -20.1 |
| $H_2O$ | $NH_3$ | CS | 0 | 4 | 0 | 2 | -21.5 | -21.2 | -21.5 | -22.4 |
| $H_2O$ | $NH_3$ | TS | 0 | 4 | 0 | 2 | -20.0 | -21.2 | -20.0 | -21.0 |
| $H_2O$ | $NH_3$ | 5+1 | 0 | 5 | 0 | 1 | -25.1 | -24.1 | -23.5 | -24.0 |
| $NH_3$ | CO | 5+1 | 1 | 0 | 0 | 5 | -0.6 | -3.0 | -2.5 | -2.0 |
| $NH_3$ | CO | CS | 2 | 0 | 0 | 4 | 5.7 | 3.7 | 5.7 | 5.6 |
| $NH_3$ | CO | TS | 2 | 0 | 0 | 4 | 4.7 | 3.7 | 4.7 | 3.7 |
| CO | $NH_3$ | FS | 3 | 0 | 0 | 3 | 13.3 | 10.4 | 14.1 | 13.3 |
| CO | $NH_3$ | MS | 3 | 0 | 0 | 3 | 10.3 | 10.4 | 9.7 | 10.3 |
| CO | $NH_3$ | CS | 4 | 0 | 0 | 2 | 22.4 | 17.0 | 22.4 | 19.0 |
| CO | $NH_3$ | TS | 4 | 0 | 0 | 2 | 14.6 | 17.0 | 14.6 | 17.0 |
| CO | $NH_3$ | 5+1 | 5 | 0 | 0 | 1 | 26.4 | 23.7 | 22.5 | 24.6 |

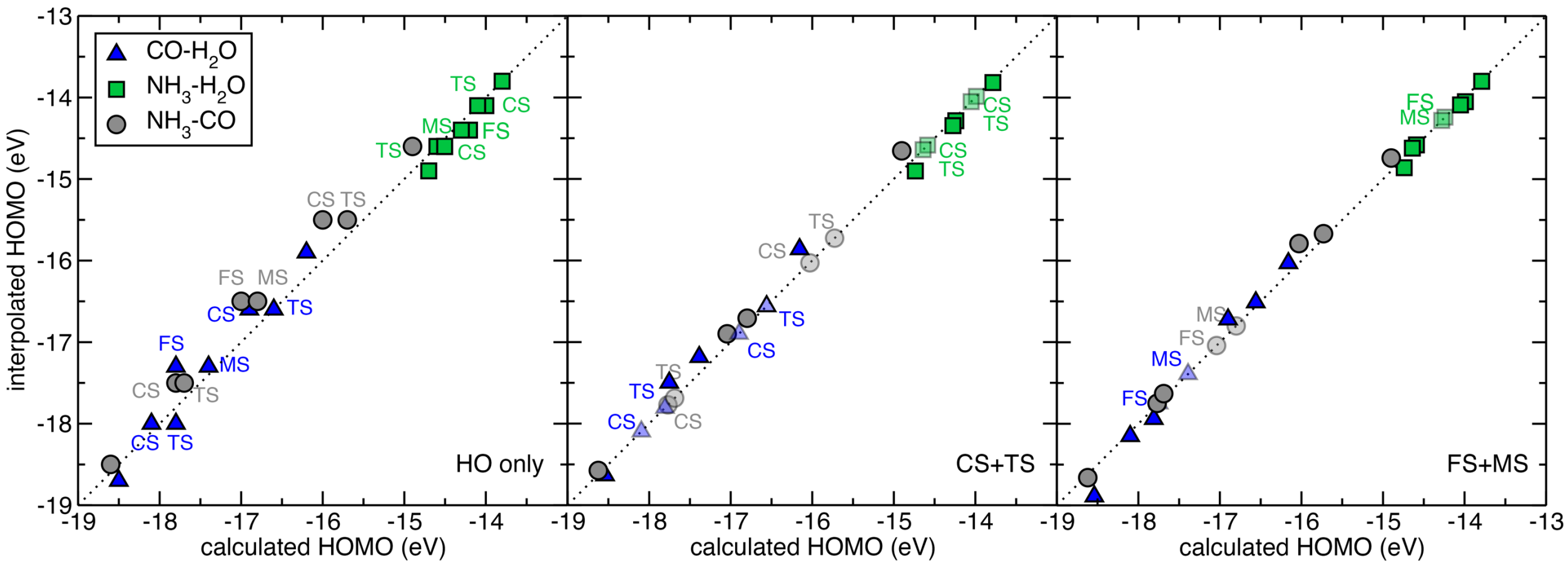

**Figure S2.** Calculated vs. linearly interpolated HOMO level (eV) for singlet Fe(II) complexes with pairs of any of the three ligands: $NH_3$, $H_2O$, and CO. From left to right: interpolation between homoleptic complexes (HO only), interpolation using homoleptic complexes as well as *cis* symmetric and *trans* symmetric complex energies (CS+TS), or interpolation using homoleptic complexes as well as *fac* and *mer* symmetric complex energies (FS+MS). Points are colored according to the pair of ligands they correspond to: $NH_3$-$H_2O$ (green squares), $NH_3$-CO (gray circles), and CO-$H_2O$ (blue triangles), as indicated in inset legend. Key isomers are annotated. Points provided for the fit are translucent, whereas the remaining points are opaque. In all panes, a black dotted parity line is shown.

**Table S9.** Mean absolute errors (MAE) in kcal/mol for Fe(II) complex $\Delta E_{H\text{-}L}$ estimates from HO-only averaging, including CS/TS in the averaging, and including FS/MS averaging for pairs of ligands, $L_1$ and $L_2$. The CS/TS and FS/MS errors are both averaged over only the points for which exact energies were not provided (interp.) as well as over all (all) points.

| | | $\Delta E_{H\text{-}L}$ MAE (kcal/mol) | | | | |
|---|---|---|---|---|---|---|
| **$L_1$** | **$L_2$** | **HO** | **CS+TS all** | **CS+TS interp.** | **FS+MS all** | **FS+MS interp.** |
| $H_2O$ | $CH_3CN$ | 6.3 | 0.7 | 1.4 | 2.6 | 3.5 |
| $CH_3CN$ | CO | 2.1 | 0.4 | 0.8 | 1.4 | 1.9 |
| $H_2O$ | CO | 5.1 | 1.2 | 2.4 | 2.3 | 3.1 |
| $NH_3$ | $H_2O$ | 1.0 | 0.6 | 1.2 | 0.8 | 1.1 |
| $NH_3$ | CO | 2.5 | 1.1 | 2.2 | 1.3 | 1.7 |
| Overall | | 3.4 | 0.8 | 1.6 | 1.7 | 2.3 |

**Table S10.** Singlet HOMO level values for pairs of two ligand types in Fe(II) complexes (in eV) along with interpolated values using only HO complexes, HO+CS/TS complexes, or HO+FS/MS complexes.

| | | | Number of ligands | | | | | Interpolated | | |
|---|---|---|---|---|---|---|---|---|---|---|
| **$L_1$** | **$L_2$** | **Symmetry** | **CO** | **$H_2O$** | **$CH_3CN$** | **$NH_3$** | **HOMO** | **HO** | **CS/TS** | **FS/MS** |
| $CH_3CN$ | $H_2O$ | 5+1 | 0 | 1 | 5 | 0 | -13.56 | -13.65 | -13.57 | -13.63 |
| $CH_3CN$ | $H_2O$ | CS | 0 | 2 | 4 | 0 | -13.82 | -13.95 | -13.82 | -13.88 |
| $CH_3CN$ | $H_2O$ | TS | 0 | 2 | 4 | 0 | -13.80 | -13.95 | -13.80 | -13.85 |
| $H_2O$ | $CH_3CN$ | FS | 0 | 3 | 3 | 0 | -14.19 | -14.25 | -14.15 | -14.19 |
| $H_2O$ | $CH_3CN$ | MS | 0 | 3 | 3 | 0 | -14.08 | -14.25 | -14.11 | -14.08 |
| $H_2O$ | $CH_3CN$ | CS | 0 | 4 | 2 | 0 | -14.48 | -14.56 | -14.48 | -14.49 |
| $H_2O$ | $CH_3CN$ | TS | 0 | 4 | 2 | 0 | -14.41 | -14.56 | -14.41 | -14.46 |
| $H_2O$ | $CH_3CN$ | 5+1 | 0 | 5 | 1 | 0 | -14.82 | -14.86 | -14.79 | -14.84 |
| | | | | | | | | | | |
| CO | $CH_3CN$ | 5+1 | 5 | 0 | 1 | 0 | -17.80 | -18.44 | -18.10 | -18.32 |
| CO | $CH_3CN$ | CS | 4 | 0 | 2 | 0 | -16.82 | -17.42 | -16.82 | -17.10 |
| CO | $CH_3CN$ | TS | 4 | 0 | 2 | 0 | -16.75 | -17.42 | -16.75 | -17.02 |
| CO | $CH_3CN$ | FS | 3 | 0 | 3 | 0 | -16.02 | -16.40 | -15.92 | -16.02 |
| CO | $CH_3CN$ | MS | 3 | 0 | 3 | 0 | -15.82 | -16.40 | -15.80 | -15.82 |
| $CH_3CN$ | CO | CS | 2 | 0 | 4 | 0 | -15.02 | -15.38 | -15.02 | -15.06 |
| $CH_3CN$ | CO | TS | 2 | 0 | 4 | 0 | -14.86 | -15.38 | -14.86 | -14.98 |
| $CH_3CN$ | CO | 5+1 | 1 | 0 | 5 | 0 | -14.11 | -14.37 | -14.10 | -14.24 |
| | | | | | | | | | | |
| CO | $H_2O$ | 5+1 | 5 | 1 | 0 | 0 | -18.54 | -18.75 | -18.63 | -18.89 |
| CO | $H_2O$ | CS | 4 | 2 | 0 | 0 | -18.10 | -18.03 | -18.10 | -18.15 |
| CO | $H_2O$ | TS | 4 | 2 | 0 | 0 | -17.81 | -18.03 | -17.81 | -17.94 |
| $H_2O$ | CO | FS | 3 | 3 | 0 | 0 | -17.75 | -17.31 | -17.50 | -17.75 |
| $H_2O$ | CO | MS | 3 | 3 | 0 | 0 | -17.39 | -17.31 | -17.18 | -17.39 |
| $H_2O$ | CO | CS | 2 | 4 | 0 | 0 | -16.90 | -16.60 | -16.90 | -16.72 |
| $H_2O$ | CO | TS | 2 | 4 | 0 | 0 | -16.56 | -16.60 | -16.56 | -16.51 |
| $H_2O$ | CO | 5+1 | 1 | 5 | 0 | 0 | -16.16 | -15.88 | -15.86 | -16.03 |
| | | | | | | | | | | |
| $H_2O$ | $NH_3$ | 5+1 | 0 | 5 | 0 | 1 | -14.74 | -14.90 | -14.90 | -14.86 |
| $H_2O$ | $NH_3$ | CS | 0 | 4 | 0 | 2 | -14.59 | -14.64 | -14.59 | -14.58 |
| $H_2O$ | $NH_3$ | TS | 0 | 4 | 0 | 2 | -14.64 | -14.64 | -14.64 | -14.62 |
| $H_2O$ | $NH_3$ | FS | 0 | 3 | 0 | 3 | -14.24 | -14.37 | -14.29 | -14.24 |
| $H_2O$ | $NH_3$ | MS | 0 | 3 | 0 | 3 | -14.28 | -14.37 | -14.35 | -14.28 |
| $NH_3$ | $H_2O$ | CS | 0 | 2 | 0 | 4 | -13.99 | -14.11 | -13.99 | -14.05 |
| $NH_3$ | $H_2O$ | TS | 0 | 2 | 0 | 4 | -14.05 | -14.11 | -14.05 | -14.09 |
| $NH_3$ | $H_2O$ | 5+1 | 0 | 1 | 0 | 5 | -13.79 | -13.85 | -13.82 | -13.80 |
| | | | | | | | | | | |
| CO | $NH_3$ | 5+1 | 5 | 0 | 0 | 1 | -18.62 | -18.48 | -18.58 | -18.66 |
| CO | $NH_3$ | CS | 4 | 0 | 0 | 2 | -17.77 | -17.50 | -17.77 | -17.75 |
| CO | $NH_3$ | TS | 4 | 0 | 0 | 2 | -17.69 | -17.50 | -17.69 | -17.63 |
| CO | $NH_3$ | FS | 3 | 0 | 0 | 3 | -17.04 | -16.52 | -16.90 | -17.04 |
| CO | $NH_3$ | MS | 3 | 0 | 0 | 3 | -16.80 | -16.52 | -16.71 | -16.80 |
| $NH_3$ | CO | CS | 2 | 0 | 0 | 4 | -16.03 | -15.54 | -16.03 | -15.79 |
| $NH_3$ | CO | TS | 2 | 0 | 0 | 4 | -15.73 | -15.54 | -15.73 | -15.67 |
| $NH_3$ | CO | 5+1 | 1 | 0 | 0 | 5 | -14.90 | -14.56 | -14.65 | -14.74 |

**Table S11.** Mean absolute errors (MAE) in eV for singlet complex HOMO levels from HO-only averaging, including CS/TS in the averaging, and including FS/MS averaging for pairs of ligands, $L_1$ and $L_2$. The CS/TS and FS/MS errors are both averaged over only the points for which exact energies were not provided (interp.) as well as over all (all) points.

| | | HOMO MAE (eV) | | | | |
|---|---|---|---|---|---|---|
| **$L_1$** | **$L_2$** | **HO** | **CS+TS all** | **CS+TS interp.** | **FS+MS all** | **FS+MS interp.** |
| $H_2O$ | $CH_3CN$ | 0.11 | 0.01 | 0.02 | 0.03 | 0.04 |
| $CH_3CN$ | CO | 0.51 | 0.05 | 0.10 | 0.17 | 0.23 |
| $H_2O$ | CO | 0.20 | 0.11 | 0.22 | 0.11 | 0.15 |
| $NH_3$ | $H_2O$ | 0.09 | 0.04 | 0.08 | 0.03 | 0.04 |
| $NH_3$ | CO | 0.30 | 0.25 | 0.50 | 0.07 | 0.09 |
| Overall | | 0.24 | 0.09 | 0.18 | 0.08 | 0.11 |

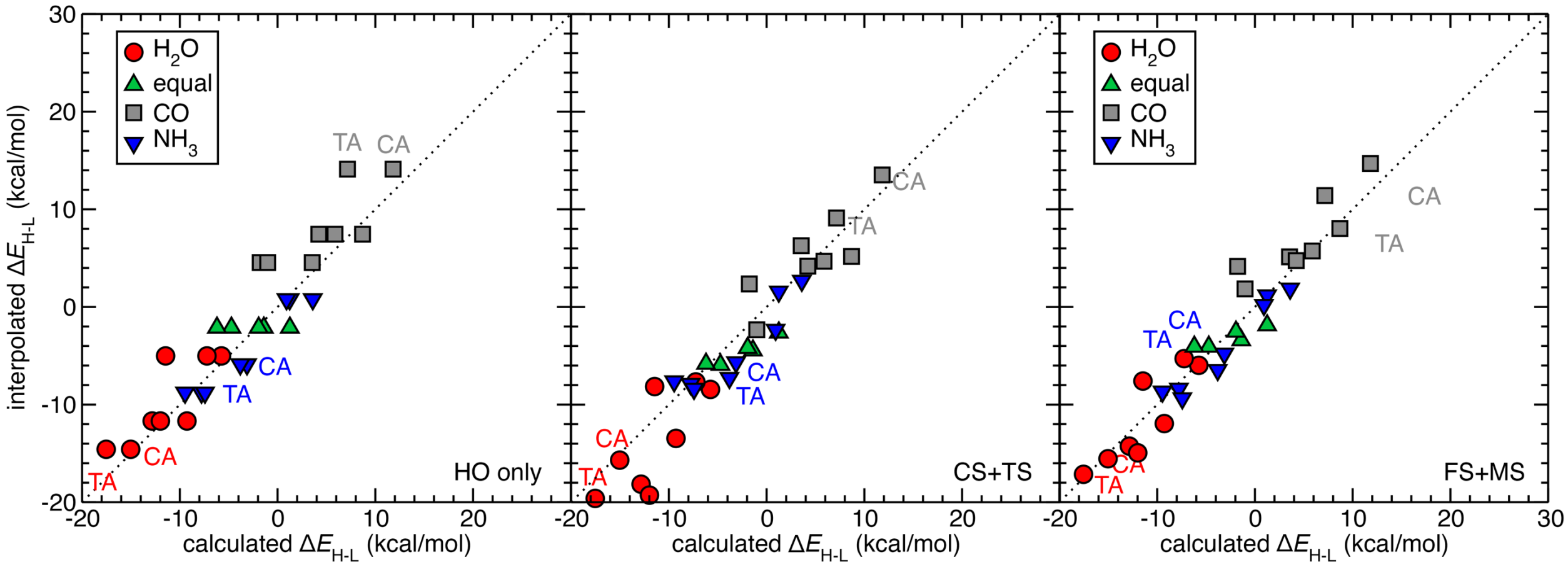

**Figure S3.** Calculated vs. linearly interpolated $\Delta E_{\text{H-L}}$ (kcal/mol) for Fe(II) complexes with at least one each of three ligands: $NH_3$, $H_2O$, and CO. From left to right: interpolation from homoleptic complexes (HO only), interpolation using homoleptic complexes as well as *cis* symmetric and *trans* symmetric complex energies derived from pairs of ligands (CS+TS), or interpolation using homoleptic complexes as well as *fac* and *mer* symmetric complex energies derived from pairs of ligands (FS+MS). Points are colored according to the ligand with the highest stoichiometric coefficient: $H_2O$ (red circles), CO (gray squares), and $NH_3$ (blue down triangles), or equal weight of all ligands (green up triangles), as indicated in inset legend. Key isomers are annotated. In all panes, a black dotted parity line is shown.

**Table S12.** $\Delta E_{H\text{-}L}$ values for Fe(II) complexes (in kcal/mol) with a combination of three ligand types, $H_2O$, CO, and $CH_3CN$, in Fe(II) complexes along with interpolated values using only HO complexes, HO+CS/TS complexes, or HO+FS/MS complexes. For the three EA isomers, $L_3$ is the ligand in the *trans* position as indicated in the main text.

| | | | | Number of ligands | | | | Interpolated | | |
|---|---|---|---|---|---|---|---|---|---|---|
| **$L_1$** | **$L_2$** | **$L_3$** | **Symmetry** | **$CH_3CN$** | **CO** | **$H_2O$** | **splitting** | **HO** | **CS/TS** | **FS/MS** |
| $H_2O$ | CO | $CH_3CN$ | TA | 1 | 1 | 4 | -15.7 | -5.7 | -16.1 | -11.5 |
| $H_2O$ | CO | $CH_3CN$ | CA | 1 | 1 | 4 | -7.7 | -5.7 | -8.8 | -7.0 |
| CO | $H_2O$ | $CH_3CN$ | FA | 1 | 2 | 3 | 3.2 | 3.9 | -1.6 | 2.5 |
| CO | $H_2O$ | $CH_3CN$ | MAC | 1 | 2 | 3 | -4.6 | 3.9 | -4.1 | 0.4 |
| CO | $H_2O$ | $CH_3CN$ | MAT | 1 | 2 | 3 | -5.5 | 3.9 | -4.6 | -1.9 |
| $CH_3CN$ | $H_2O$ | CO | FA | 2 | 1 | 3 | 5.2 | 6.1 | 1.8 | 4.7 |
| $CH_3CN$ | $H_2O$ | CO | MAC | 2 | 1 | 3 | -2.3 | 6.1 | -2.9 | 2.4 |
| $CH_3CN$ | $H_2O$ | CO | MAT | 2 | 1 | 3 | -1.7 | 6.1 | -4.4 | 0.3 |
| | | | | | | | | | | |
| CO | $CH_3CN$ | $H_2O$ | EA | 2 | 2 | 2 | 6.1 | 15.7 | 10.0 | 12.1 |
| $H_2O$ | CO | $CH_3CN$ | DTS | 2 | 2 | 2 | 6.2 | 15.7 | 10.4 | 12.1 |
| $H_2O$ | CO | $CH_3CN$ | DCS | 2 | 2 | 2 | 10.9 | 15.7 | 13.1 | 15.1 |
| $H_2O$ | $CH_3CN$ | CO | EA | 2 | 2 | 2 | 12.0 | 15.7 | 11.7 | 13.5 |
| CO | $H_2O$ | $CH_3CN$ | EA | 2 | 2 | 2 | 12.9 | 15.7 | 12.1 | 13.6 |
| | | | | | | | | | | |
| $H_2O$ | CO | $CH_3CN$ | FA | 1 | 3 | 2 | 9.0 | 13.4 | 14.5 | 13.5 |
| $H_2O$ | CO | $CH_3CN$ | MAC | 1 | 3 | 2 | 8.4 | 13.4 | 14.6 | 13.4 |
| $H_2O$ | CO | $CH_3CN$ | MAT | 1 | 3 | 2 | 3.5 | 13.4 | 9.9 | 11.1 |
| $CH_3CN$ | CO | $H_2O$ | FA | 2 | 3 | 1 | 22.9 | 25.2 | 21.9 | 25.3 |
| $CH_3CN$ | CO | $H_2O$ | MAC | 2 | 3 | 1 | 21.0 | 25.2 | 21.4 | 23.0 |
| $CH_3CN$ | CO | $H_2O$ | MAT | 2 | 3 | 1 | 23.1 | 25.2 | 22.4 | 22.9 |
| CO | $H_2O$ | $CH_3CN$ | TA | 1 | 4 | 1 | 15.7 | 23.0 | 21.4 | 20.7 |
| CO | $H_2O$ | $CH_3CN$ | CA | 1 | 4 | 1 | 19.4 | 23.0 | 21.7 | 23.1 |
| | | | | | | | | | | |
| $H_2O$ | $CH_3CN$ | CO | FA | 3 | 1 | 2 | 12.1 | 17.9 | 15.7 | 17.4 |
| $H_2O$ | $CH_3CN$ | CO | MAC | 3 | 1 | 2 | 13.8 | 17.9 | 16.7 | 17.3 |
| $H_2O$ | $CH_3CN$ | CO | MAT | 3 | 1 | 2 | 9.7 | 17.9 | 14.2 | 15.2 |
| CO | $CH_3CN$ | $H_2O$ | FA | 3 | 2 | 1 | 23.6 | 27.4 | 26.1 | 26.9 |
| CO | $CH_3CN$ | $H_2O$ | MAC | 3 | 2 | 1 | 25.2 | 27.4 | 24.7 | 24.8 |
| CO | $CH_3CN$ | $H_2O$ | MAT | 3 | 2 | 1 | 23.8 | 27.4 | 24.8 | 24.8 |
| $CH_3CN$ | $H_2O$ | CO | CA | 4 | 1 | 1 | 25.2 | 29.7 | 26.3 | 29.2 |
| $CH_3CN$ | CO | $H_2O$ | TA | 4 | 1 | 1 | 26.9 | 29.7 | 25.9 | 27.0 |

**Table S13.** $\Delta E_{H\text{-}L}$ values for Fe(II) complexes (in kcal/mol) with a combination of three ligand types, $H_2O$, CO, and $NH_3$, in Fe(II) complexes along with interpolated values using only HO complexes, HO+CS/TS complexes, or HO+FS/MS complexes. For the three EA isomers, $L_3$ is the ligand in the *trans* position as indicated in the main text.

| | | | | **Number of ligands** | | | | **Interpolated** | | |
|---|---|---|---|---|---|---|---|---|---|---|
| **$L_1$** | **$L_2$** | **$L_3$** | **symmetry** | **$NH_3$** | **CO** | **$H_2O$** | **splitting** | **HO** | **CS/TS** | **FS/MS** |
| $H_2O$ | CO | $NH_3$ | CA | 1 | 1 | 4 | -15.0 | -14.6 | -15.7 | -15.5 |
| $H_2O$ | CO | $NH_3$ | TA | 1 | 1 | 4 | -17.5 | -14.6 | -19.6 | -17.2 |
| CO | $H_2O$ | $NH_3$ | FA | 1 | 2 | 3 | -5.7 | -5.0 | -8.4 | -6.0 |
| CO | $H_2O$ | $NH_3$ | MAC | 1 | 2 | 3 | -7.2 | -5.0 | -7.7 | -5.3 |
| CO | $H_2O$ | $NH_3$ | MAT | 1 | 2 | 3 | -11.5 | -5.0 | -8.2 | -7.6 |
| $NH_3$ | $H_2O$ | CO | FA | 2 | 1 | 3 | -9.3 | -11.7 | -13.5 | -12.0 |
| $NH_3$ | $H_2O$ | CO | MAC | 2 | 1 | 3 | -12.8 | -11.7 | -18.2 | -14.3 |
| $NH_3$ | $H_2O$ | CO | MAT | 2 | 1 | 3 | -12.0 | -11.7 | -19.3 | -14.9 |
| | | | | | | | | | | |
| $H_2O$ | CO | $NH_3$ | DCS | 2 | 2 | 2 | 1.3 | -2.1 | -2.6 | -1.9 |
| H2O | CO | $NH_3$ | DTS | 2 | 2 | 2 | -4.7 | -2.1 | -5.9 | -4.1 |
| CO | $NH_3$ | $H_2O$ | EA | 2 | 2 | 2 | -1.4 | -2.1 | -4.5 | -3.4 |
| H2O | $NH_3$ | CO | EA | 2 | 2 | 2 | -6.2 | -2.1 | -5.8 | -4.1 |
| CO | $H_2O$ | $NH_3$ | EA | 2 | 2 | 2 | -1.9 | -2.1 | -4.2 | -2.5 |
| | | | | | | | | | | |
| H2O | CO | $NH_3$ | FA | 1 | 3 | 2 | 3.5 | 4.5 | 6.3 | 5.1 |
| H2O | CO | $NH_3$ | MAC | 1 | 3 | 2 | -1.8 | 4.5 | 2.4 | 4.1 |
| H2O | CO | $NH_3$ | MAT | 1 | 3 | 2 | -1.0 | 4.5 | -2.3 | 1.8 |
| NH3 | CO | $H_2O$ | FA | 2 | 3 | 1 | 8.7 | 7.4 | 5.2 | 8.0 |
| NH3 | CO | $H_2O$ | MAC | 2 | 3 | 1 | 5.9 | 7.4 | 4.7 | 5.7 |
| NH3 | CO | $H_2O$ | MAT | 2 | 3 | 1 | 4.2 | 7.4 | 4.2 | 4.7 |
| CO | $H_2O$ | $NH_3$ | CA | 1 | 4 | 1 | 11.8 | 14.1 | 13.5 | 14.7 |
| CO | $H_2O$ | $NH_3$ | TA | 1 | 4 | 1 | 7.1 | 14.1 | 9.1 | 11.4 |
| | | | | | | | | | | |
| H2O | $NH_3$ | CO | FA | 3 | 1 | 2 | -7.8 | -8.8 | -7.9 | -8.4 |
| H2O | $NH_3$ | CO | MAC | 3 | 1 | 2 | -7.4 | -8.8 | -8.4 | -9.4 |
| H2O | $NH_3$ | CO | MAT | 3 | 1 | 2 | -9.5 | -8.8 | -7.6 | -8.7 |
| CO | $NH_3$ | $H_2O$ | FA | 3 | 2 | 1 | 3.6 | 0.8 | 2.7 | 1.9 |
| CO | $NH_3$ | $H_2O$ | MAC | 3 | 2 | 1 | 1.2 | 0.8 | 1.6 | 1.2 |
| CO | $NH_3$ | $H_2O$ | MAT | 3 | 2 | 1 | 0.9 | 0.8 | -2.3 | 0.2 |
| $NH_3$ | $H_2O$ | CO | CA | 4 | 1 | 1 | -3.1 | -5.9 | -5.7 | -4.8 |
| $NH_3$ | CO | $H_2O$ | TA | 4 | 1 | 1 | -3.8 | -5.9 | -7.3 | -6.5 |

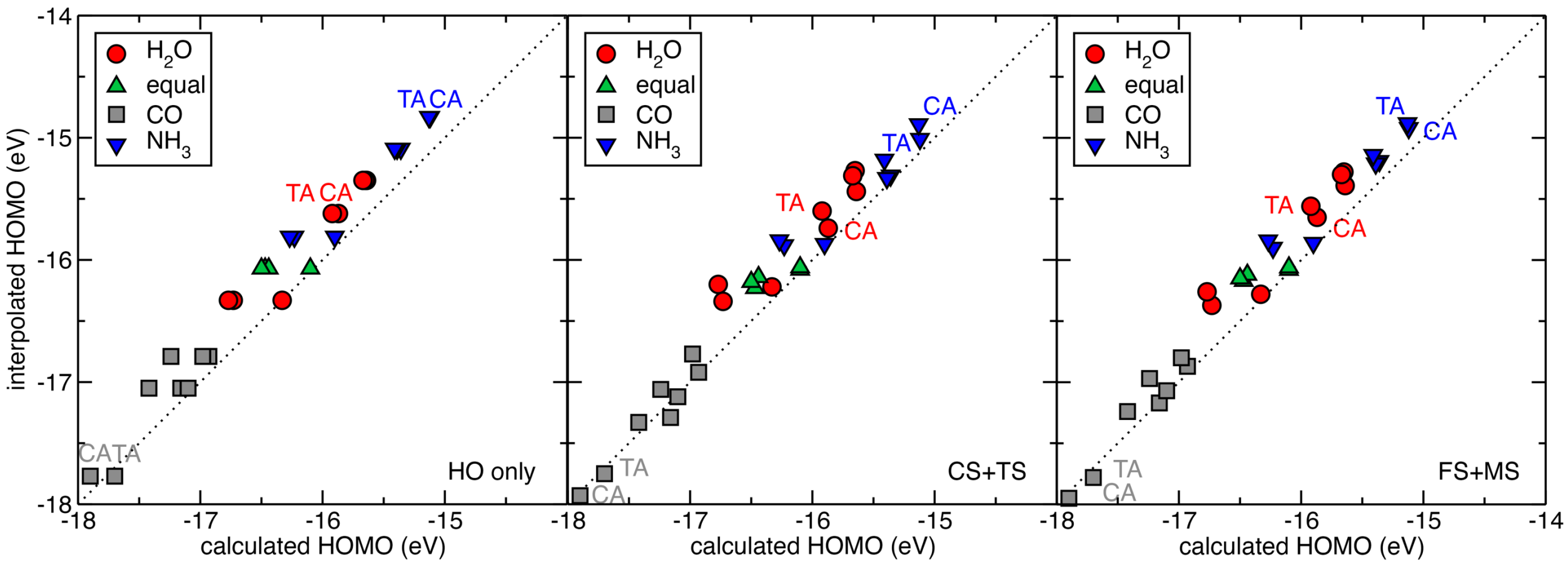

**Figure S4.** Calculated vs. linearly interpolated HOMO level (eV) for singlet Fe(II) complexes with at least one each of three ligands: $NH_3$, $H_2O$, and CO. From left to right: interpolation from homoleptic complexes (HO only), interpolation using homoleptic complexes as well as *cis* symmetric and *trans* symmetric complex energies derived from pairs of ligands (CS+TS), or interpolation using homoleptic complexes as well as *fac* and *mer* symmetric complex energies derived from pairs of ligands (FS+MS). Points are colored according to the ligand with the highest stoichiometric coefficient: $H_2O$ (red circles), CO (gray squares), and $NH_3$ (blue down triangles), or equal weight of all ligands (green up triangles), as indicated in inset legend. Key isomers are annotated. In all panes, a black dotted parity line is shown.

**Table S14.** Singlet HOMO level values with a combination of three ligand types, $H_2O$, CO, and $CH_3CN$, in Fe(II) complexes (in eV) along with interpolated values using only HO complexes, HO+CS/TS complexes, or HO+FS/MS complexes. For the three EA isomers, $L_3$ is the ligand in the *trans* position as indicated in the main text.

| | | | | Number of ligands | | | | Interpolated | | |
|---|---|---|---|---|---|---|---|---|---|---|
| **$L_1$** | **$L_2$** | **$L_3$** | **symmetry** | **$CH_3CN$** | **CO** | **$H_2O$** | **HOMO** | **HO** | **CS/TS** | **FS/MS** |
| $H_2O$ | CO | $CH_3CN$ | CA | 1 | 1 | 4 | -15.71 | -15.58 | -15.54 | -15.60 |
| $H_2O$ | CO | $CH_3CN$ | TA | 1 | 1 | 4 | -15.50 | -15.58 | -15.69 | -15.48 |
| CO | $H_2O$ | $CH_3CN$ | FA | 1 | 2 | 3 | -16.40 | -16.29 | -16.14 | -16.32 |
| CO | $H_2O$ | $CH_3CN$ | MAC | 1 | 2 | 3 | -16.34 | -16.29 | -16.46 | -16.31 |
| CO | $H_2O$ | $CH_3CN$ | MAT | 1 | 2 | 3 | -16.13 | -16.29 | -16.31 | -16.20 |
| $CH_3CN$ | $H_2O$ | CO | FA | 2 | 1 | 3 | -15.23 | -15.27 | -15.23 | -15.30 |
| $CH_3CN$ | $H_2O$ | CO | MAC | 2 | 1 | 3 | -15.13 | -15.27 | -15.06 | -15.20 |
| $CH_3CN$ | $H_2O$ | CO | MAT | 2 | 1 | 3 | -15.30 | -15.27 | -15.18 | -15.18 |
| | | | | | | | | | | |
| $H_2O$ | CO | $CH_3CN$ | DCS | 2 | 2 | 2 | -15.89 | -15.99 | -15.76 | -15.90 |
| $H_2O$ | CO | $CH_3CN$ | DTS | 2 | 2 | 2 | -15.75 | -15.99 | -15.77 | -15.80 |
| CO | $CH_3CN$ | $H_2O$ | EA | 2 | 2 | 2 | -15.76 | -15.99 | -15.80 | -15.82 |
| $H_2O$ | $CH_3CN$ | CO | EA | 2 | 2 | 2 | -15.65 | -15.99 | -15.62 | -15.81 |
| CO | $H_2O$ | $CH_3CN$ | EA | 2 | 2 | 2 | -15.87 | -15.99 | -15.87 | -15.87 |
| | | | | | | | | | | |
| $H_2O$ | CO | $CH_3CN$ | FA | 1 | 3 | 2 | -16.96 | -17.01 | -16.86 | -16.91 |
| $H_2O$ | CO | $CH_3CN$ | MAC | 1 | 3 | 2 | -16.77 | -17.01 | -16.82 | -16.87 |
| $H_2O$ | CO | $CH_3CN$ | MAT | 1 | 3 | 2 | -16.76 | -17.01 | -16.65 | -16.77 |
| $CH_3CN$ | CO | $H_2O$ | FA | 2 | 3 | 1 | -16.43 | -16.71 | -16.56 | -16.61 |
| $CH_3CN$ | CO | $H_2O$ | MAC | 2 | 3 | 1 | -16.29 | -16.71 | -16.41 | -16.50 |
| $CH_3CN$ | CO | $H_2O$ | MAT | 2 | 3 | 1 | -16.35 | -16.71 | -16.33 | -16.46 |
| CO | $H_2O$ | $CH_3CN$ | CA | 1 | 4 | 1 | -17.46 | -17.73 | -17.46 | -17.63 |
| CO | $H_2O$ | $CH_3CN$ | TA | 1 | 4 | 1 | -17.14 | -17.73 | -17.28 | -17.48 |
| | | | | | | | | | | |
| $H_2O$ | $CH_3CN$ | CO | FA | 3 | 1 | 2 | -14.77 | -14.97 | -14.60 | -14.77 |
| $H_2O$ | $CH_3CN$ | CO | MAC | 3 | 1 | 2 | -14.85 | -14.97 | -14.92 | -14.76 |
| $H_2O$ | $CH_3CN$ | CO | MAT | 3 | 1 | 2 | -14.80 | -14.97 | -14.84 | -14.72 |
| CO | $CH_3CN$ | $H_2O$ | FA | 3 | 2 | 1 | -15.38 | -15.69 | -15.19 | -15.49 |
| CO | $CH_3CN$ | $H_2O$ | MAC | 3 | 2 | 1 | -15.38 | -15.69 | -15.15 | -15.45 |
| CO | $CH_3CN$ | $H_2O$ | MAT | 3 | 2 | 1 | -15.27 | -15.69 | -15.28 | -15.44 |
| $CH_3CN$ | $H_2O$ | CO | CA | 4 | 1 | 1 | -14.45 | -14.67 | -14.29 | -14.47 |
| $CH_3CN$ | CO | $H_2O$ | TA | 4 | 1 | 1 | -14.42 | -14.67 | -14.33 | -14.42 |

**Table S15.** Singlet HOMO level values with a combination of three ligand types, $H_2O$, CO, and $NH_3$, in Fe(II) complexes (in eV) along with interpolated values using only HO complexes, HO+CS/TS complexes, or HO+FS/MS complexes. For the three EA isomers, $L_3$ is the ligand in the *trans* position as indicated in the main text.

| | | | | **Number of ligands** | | | | **Interpolated** | | |
|---|---|---|---|---|---|---|---|---|---|---|
| **$L_1$** | **$L_2$** | **$L_3$** | **Symmetry** | **$NH_3$** | **CO** | **$H_2O$** | **HOMO** | **HO** | **CS/TS** | **FS/MS** |
| $H_2O$ | CO | $NH_3$ | CA | 1 | 1 | 4 | -15.87 | -15.62 | -15.74 | -15.65 |
| $H_2O$ | CO | $NH_3$ | TA | 1 | 1 | 4 | -15.92 | -15.62 | -15.60 | -15.56 |
| $NH_3$ | $H_2O$ | CO | FA | 2 | 1 | 3 | -15.64 | -15.35 | -15.44 | -15.39 |
| $NH_3$ | $H_2O$ | CO | MAC | 2 | 1 | 3 | -15.65 | -15.35 | -15.27 | -15.28 |
| $NH_3$ | $H_2O$ | CO | MAT | 2 | 1 | 3 | -15.67 | -15.35 | -15.31 | -15.30 |
| CO | $H_2O$ | $NH_3$ | FA | 1 | 2 | 3 | -16.73 | -16.33 | -16.34 | -16.37 |
| CO | $H_2O$ | $NH_3$ | MAC | 1 | 2 | 3 | -16.77 | -16.33 | -16.20 | -16.26 |
| CO | $H_2O$ | $NH_3$ | MAT | 1 | 2 | 3 | -16.33 | -16.33 | -16.22 | -16.28 |
| | | | | | | | | | | |
| $H_2O$ | CO | $NH_3$ | DCS | 2 | 2 | 2 | -16.47 | -16.07 | -16.23 | -16.17 |
| $H_2O$ | CO | $NH_3$ | DTS | 2 | 2 | 2 | -16.10 | -16.07 | -16.08 | -16.08 |
| CO | $NH_3$ | $H_2O$ | EA | 2 | 2 | 2 | -16.44 | -16.07 | -16.14 | -16.12 |
| $H_2O$ | $NH_3$ | CO | EA | 2 | 2 | 2 | -16.10 | -16.07 | -16.06 | -16.06 |
| CO | $H_2O$ | $NH_3$ | EA | 2 | 2 | 2 | -16.50 | -16.07 | -16.18 | -16.15 |
| | | | | | | | | | | |
| $H_2O$ | CO | $NH_3$ | FA | 1 | 3 | 2 | -17.42 | -17.05 | -17.33 | -17.24 |
| $H_2O$ | CO | $NH_3$ | MAC | 1 | 3 | 2 | -17.16 | -17.05 | -17.29 | -17.17 |
| $H_2O$ | CO | $NH_3$ | MAT | 1 | 3 | 2 | -17.10 | -17.05 | -17.12 | -17.07 |
| $NH_3$ | CO | $H_2O$ | FA | 2 | 3 | 1 | -17.24 | -16.79 | -17.06 | -16.97 |
| $NH_3$ | CO | $H_2O$ | MAC | 2 | 3 | 1 | -16.93 | -16.79 | -16.92 | -16.87 |
| $NH_3$ | CO | $H_2O$ | MAT | 2 | 3 | 1 | -16.98 | -16.79 | -16.77 | -16.80 |
| CO | $H_2O$ | $NH_3$ | CA | 1 | 4 | 1 | -17.90 | -17.77 | -17.93 | -17.95 |
| CO | $H_2O$ | $NH_3$ | TA | 1 | 4 | 1 | -17.70 | -17.77 | -17.75 | -17.78 |
| | | | | | | | | | | |
| $H_2O$ | $NH_3$ | CO | FA | 3 | 1 | 2 | -15.36 | -15.09 | -15.31 | -15.19 |
| $H_2O$ | $NH_3$ | CO | MAC | 3 | 1 | 2 | -15.39 | -15.09 | -15.33 | -15.21 |
| $H_2O$ | $NH_3$ | CO | MAT | 3 | 1 | 2 | -15.41 | -15.09 | -15.18 | -15.14 |
| CO | $NH_3$ | $H_2O$ | FA | 3 | 2 | 1 | -16.23 | -15.81 | -15.88 | -15.90 |
| CO | $NH_3$ | $H_2O$ | MAC | 3 | 2 | 1 | -16.27 | -15.81 | -15.84 | -15.84 |
| CO | $NH_3$ | $H_2O$ | MAT | 3 | 2 | 1 | -15.90 | -15.81 | -15.87 | -15.86 |
| $NH_3$ | $H_2O$ | CO | CA | 4 | 1 | 1 | -15.12 | -14.83 | -15.01 | -14.92 |
| $NH_3$ | CO | $H_2O$ | TA | 4 | 1 | 1 | -15.13 | -14.83 | -14.89 | -14.88 |

**Text S3.** Expressions for interpolating ternary complex energies from CS/TS binary complexes. We indicate the stoichiometry in the complex to distinguish where the energies are derived from:

$$E(\text{CA-}(L_1)_4(L_2)_1(L_3)_1) = ½(E(\text{CS }(L_1)_4(L_2)_2)+E(\text{CS }(L_1)_4(L_3)_2))$$

$$E(\text{TA-}(L_1)_4(L_2)_1(L_3)_1) = ½(E(\text{TS }(L_1)_4(L_2)_2)+E(\text{TS }(L_1)_4(L_3)_2))$$

$$E(\text{FA-}(L_1)_3(L_2)_2(L_3)_1) = ½(E(\text{CS }(L_1)_4(L_3)_2)+E(\text{CS }(L_2)_4(L_1)_2))$$

$$E(\text{MAC-}(L_1)_3(L_2)_2(L_3)_1) = ½(E(\text{TS }(L_1)_4(L_3)_2)+E(\text{CS }(L_2)_4(L_1)_2))$$

$$E(\text{MAT-}(L_1)_3(L_2)_2(L_3)_1) = ½(E(\text{TS }(L_1)_4(L_3)_2)+E(\text{TS }(L_2)_4(L_1)_2))$$

$$\text{E(DCS-}(L_1)_2(L_2)_2(L_3)_2) = 1/6(E(\text{CS }(L_1)_4(L_2)_2+ E(\text{CS }(L_2)_4(L_1)_2+ E(\text{CS }(L_1)_4(L_3)_2 + E(\text{CS }(L_3)_4(L_1)_2 + E(\text{CS }(L_2)_4(L_3)_2 + E(\text{CS }(L_3)_4(L_2)_2)$$

$$\text{E(DCS-}(L_1)_2(L_2)_2(L_3)_2) = 1/6(E(\text{TS }(L_1)_4(L_2)_2+ E(\text{TS }(L_2)_4(L_1)_2+ E(\text{TS }(L_1)_4(L_3)_2 + E(\text{TS }(L_3)_4(L_1)_2 + E(\text{TS }(L_2)_4(L_3)_2 + E(\text{TS }(L_3)_4(L_2)_2)$$

$$\text{E(EA-}(L_1)_2(L_2)_2(L_3)_2) = 1/6(E(\text{CS }(L_1)_4(L_2)_2+ E(\text{CS }(L_2)_4(L_1)_2+ E(\text{TS }(L_1)_4(L_3)_2 + E(\text{TS }(L_3)_4(L_1)_2 + E(\text{TS }(L_2)_4(L_3)_2 + E(\text{TS }(L_3)_4(L_2)_2)$$

where the third ligand in the EA complex is the one that is *trans* to itself so energies involving that ligand are derived from the binary TS complexes, whereas the remainder are derived from CS complexes.

**Table S16.** Mean absolute error (MAE) for singlet complex HOMO levels (in eV) and Fe(II) complex $\Delta E_{\text{H-L}}$ (in kcal/mol) for combinations of ligands, $L_1$, $L_2$, and $L_3$ evaluated with three averaging schemes: HO-only, CS/TS-derived averaging, and FS/MS-derived averaging.

| | | | **HOMO MAE (eV)** | | | **$\Delta E_{\text{H-L}}$ MAE (kcal/mol)** | | |
|---|---|---|---|---|---|---|---|---|
| **$L_1$** | **$L_2$** | **$L_3$** | **HO** | **CS+TS** | **FS+MS** | **HO** | **CS+TS** | **FS+MS** |
| $H_2O$ | CO | $NH_3$ | 0.26 | 0.19 | 0.22 | 2.31 | 2.30 | 1.78 |
| $H_2O$ | CO | $CH_3CN$ | 0.21 | 0.10 | 0.09 | 5.23 | 2.45 | 3.22 |
| Overall | | | 0.24 | 0.15 | 0.16 | 3.77 | 2.38 | 2.50 |

**Table S17.** Properties of 56 homoleptic Fe(II) complexes derived from either homoleptic examples in the CSD (36 "HO") with neutral ligands or from neutral ligands only present in binary and ternary Fe(II) complexes (20 "B or T") from DFT. Both the $\Delta E_{H\text{-}L}$ (in kcal/mol) and the HOMO level of the singlet complex (in eV) are shown. Complexes are distinguished by stoichiometry.

| # atoms | Stoichiometry | Source | refcode | $\Delta E_{H\text{-}L}$ (kcal/mol) | HOMO level (eV) |
|---|---|---|---|---|---|
| 37 | Fe1C12N6H18 | HO | ACEYOW01 | -4.98 | -12.97 |
| 19 | Fe1O6H12 | HO | AMAVOB | -26.78 | -15.11 |
| 55 | Fe1C18N12H24 | HO | AXAKIT01 | -9.06 | -10.90 |
| 97 | Fe1C30N6O6H54 | HO | BIRSAZ | -30.77 | -10.48 |
| 67 | Fe1C30N6H30 | HO | BUSVAO | -16.02 | -11.82 |
| 79 | Fe1C24O6H48 | HO | BUSVES | -33.88 | -12.51 |
| 25 | Fe1N6H18 | HO | CACDIW | -9.64 | -13.58 |
| 73 | Fe1C18N6O6H42 | HO | CALMOS01 | -21.32 | -10.50 |
| 13 | Fe1C6O6 | HO | CEHHON | 30.03 | -19.46 |
| 61 | Fe1C18N6O6H30 | HO | CEMGUX | -23.44 | -11.38 |
| 121 | Fe1C42N6O6H66 | HO | DECRAE | 45.81 | -10.93 |
| 67 | Fe1C18N18H30 | HO | DEDLAB | -8.03 | -11.66 |
| 73 | Fe1C18N6S6H42 | HO | FAVDOV | -17.25 | -10.27 |
| 85 | Fe1C18N12S6H48 | HO | FEGGAZ | -24.89 | -10.13 |
| 73 | Fe1C30N6O6H30 | HO | FEHPYO | -26.10 | -9.91 |
| 49 | Fe1C18N6O6H18 | HO | FEISXC01 | -5.02 | -12.21 |
| 73 | Fe1C24N12H36 | HO | FIWQUY | -8.06 | -10.44 |
| 169 | Fe1C84N24O12H48 | HO | GOGSAZ01 | -0.40 | -9.19 |
| 61 | Fe1C12N24H24 | HO | HIPXOW | -5.43 | -11.37 |
| 127 | Fe1C36N6O6H78 | HO | HUGVIQ | -30.16 | -10.36 |
| 97 | Fe1C24N24H48 | HO | JANSAS01 | -1.77 | -11.43 |
| 61 | Fe1C12O6S6H36 | HO | JOHCOZ | -30.25 | -10.56 |
| 61 | Fe1C12N24H24 | HO | JUVHEN | -1.95 | -11.81 |
| 37 | Fe1C12N6H18 | HO | MICYFE10 | 43.67 | -13.35 |
| 115 | Fe1C30N24H60 | HO | NIGXUY01 | -1.83 | -11.32 |
| 79 | Fe1C18N24Cl6H30 | HO | PEJQIF01 | -1.93 | -11.57 |
| 55 | Fe1C6N30H18 | HO | SIDMAW | -2.30 | -12.01 |
| 55 | Fe1C6N30H18 | HO | SIDMEA | -7.85 | -12.17 |
| 79 | Fe1C18N24H36 | HO | TUNBIN | -1.14 | -11.55 |
| 37 | Fe1C6O6H24 | HO | USIMOA | -25.75 | -13.70 |
| 97 | Fe1C24N24Cl6H42 | HO | VIFNAC01 | -2.71 | -11.82 |
| 115 | Fe1C30N24H60 | HO | VOJPAM | -1.60 | -11.35 |
| 91 | Fe1C18N42H30 | HO | XOJCIL | -1.79 | -11.14 |
| 79 | Fe1C18N24Br6H30 | HO | YAGYIP01 | -1.49 | -11.69 |
| 79 | Fe1C18N24I6H30 | HO | YAGYUB01 | -1.66 | -11.28 |
| 97 | Fe1C24N24H48 | HO | YEQKIR | -1.43 | -11.32 |
| 97 | Fe1C18O18P6H54 | B or T | -- | 6.37 | -12.08 |
| 115 | Fe1C42N24H48 | B or T | -- | -19.67 | -10.27 |
| 79 | Fe1C18P6H54 | B or T | -- | -16.94 | -12.36 |
| 61 | Fe1C12N6O12H30 | B or T | -- | -38.56 | -10.26 |
| 121 | Fe1C60N12H48 | B or T | -- | -17.54 | -10.69 |
| 145 | Fe1C72N12H60 | B or T | -- | -16.08 | -10.04 |
| 91 | Fe1C36N12O6H36 | B or T | -- | -17.00 | -11.21 |
| 151 | Fe1C36O18P6H90 | B or T | -- | 9.61 | -11.51 |
| 169 | Fe1C60O12P6H90 | B or T | -- | -10.82 | -10.68 |
| 151 | Fe1C72N24O6H48 | B or T | -- | -16.64 | -10.11 |
| 61 | Fe1C24N12Cl6H18 | B or T | -- | -16.56 | -13.24 |
| 73 | Fe1C24N18H30 | B or T | -- | -15.67 | -11.23 |
| 97 | Fe1C48N6H42 | B or T | -- | 43.43 | -10.79 |
| 139 | Fe1C54N42H42 | B or T | -- | -4.66 | -10.19 |
| 133 | Fe1C60N12O12H48 | B or T | -- | -22.25 | -9.33 |
| 73 | Fe1C12N12H48 | B or T | -- | -9.36 | -11.36 |
| 103 | Fe1C42N24H36 | B or T | -- | -5.43 | -10.10 |
| 19 | Fe1C6N6H6 | B or T | -- | 43.66 | -15.08 |
| 127 | Fe1C60N12O6H48 | B or T | -- | -25.06 | -9.85 |
| 97 | Fe1C24N24H48 | B or T | -- | -1.13 | -11.32 |

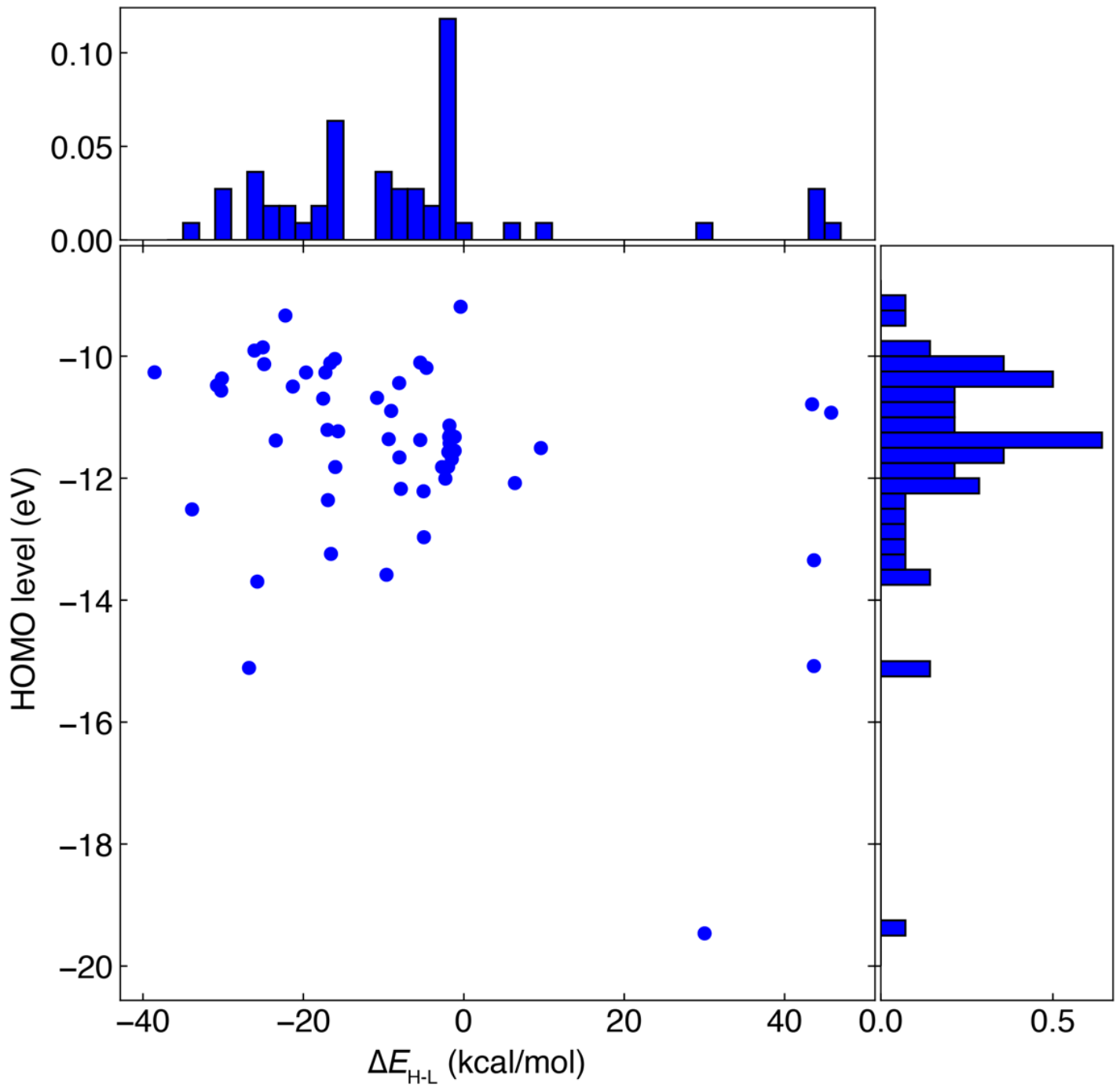

**Figure S5.** The $\Delta E_{\mathrm{H\text{-}L}}$ (in kcal/mol) vs. singlet HOMO level (in eV) for 56 homoleptic Fe(II) complexes shown as a scatter plot (middle) with normalized marginal 1D histograms at top and right, respectively.

**Table S18.** Twelve most common ligands in target zone (i.e., $\Delta E_{\mathrm{H\text{-}L}}$ from -4 to +4 kcal/mol and singlet HOMO level from -14.0 to -13.0 eV) for Fe(II) complexes and the number of times the homoleptic-only averaging predicts that a complex in the zone will have that ligand in a binary complex or both binary and ternary complexes. Refcodes in italics refer to any complex that contains that ligand, whereas the remainder are the precise homoleptic Fe(II) complexes. The top 10 most frequent ligands were selected along with two high-frequency ligands (MeCN and ClPyz) for added diversity.

| Refcode | Description | Short name | # binary | # binary or ternary | $\Delta E_{\mathrm{H\text{-}L}}$ (kcal/mol) | HOMO (eV) |
|---|---|---|---|---|---|---|
| CEHHON | CO | CO | 23 | 5071 | 30.0 | -19.46 |
| AMAVOB | H2O | H2O | 4 | 671 | -26.8 | -15.11 |
| RIFLEY | hydrogen isocyanide | CNH | 6 | 1016 | 43.7 | -15.08 |
| *GOVKEK_comp_0* | 4,4'-bipyridine | bpy20 | 2 | 364 | -17.5 | -10.69 |
| USIMOA | methanol | MeOH | 2 | 484 | -25.7 | -13.70 |
| FAVDOV | dimethylthioformamide | DMTF | 2 | 374 | -17.3 | -10.27 |
| CACDIW | ammonia | NH3 | 4 | 532 | -9.6 | -13.58 |
| *BOYWEU_comp_0* | dimethyltriazolopyrimidine | DMTP | 2 | 348 | -19.7 | -10.27 |
| *ARENUG_comp_0* | 2-chloropyrazine | ClPyz | 4 | 377 | -16.6 | -13.24 |
| *MICYFE_10* | methyl isocyanide | $CH_3CN$ (misc) | 6 | 536 | 43.7 | -13.35 |
| ACEYOW01 | NCCH3 (acetonitrile) | MeCN | 2 | 439 | -5.0 | -12.97 |
| *AMUZEP_comp_0* | 4-(5-(pyridine-4-yl)-1,3,4-oxadiazol-2-yl pyridine | bpy25 | 2 | 369 | -16.6 | -10.11 |

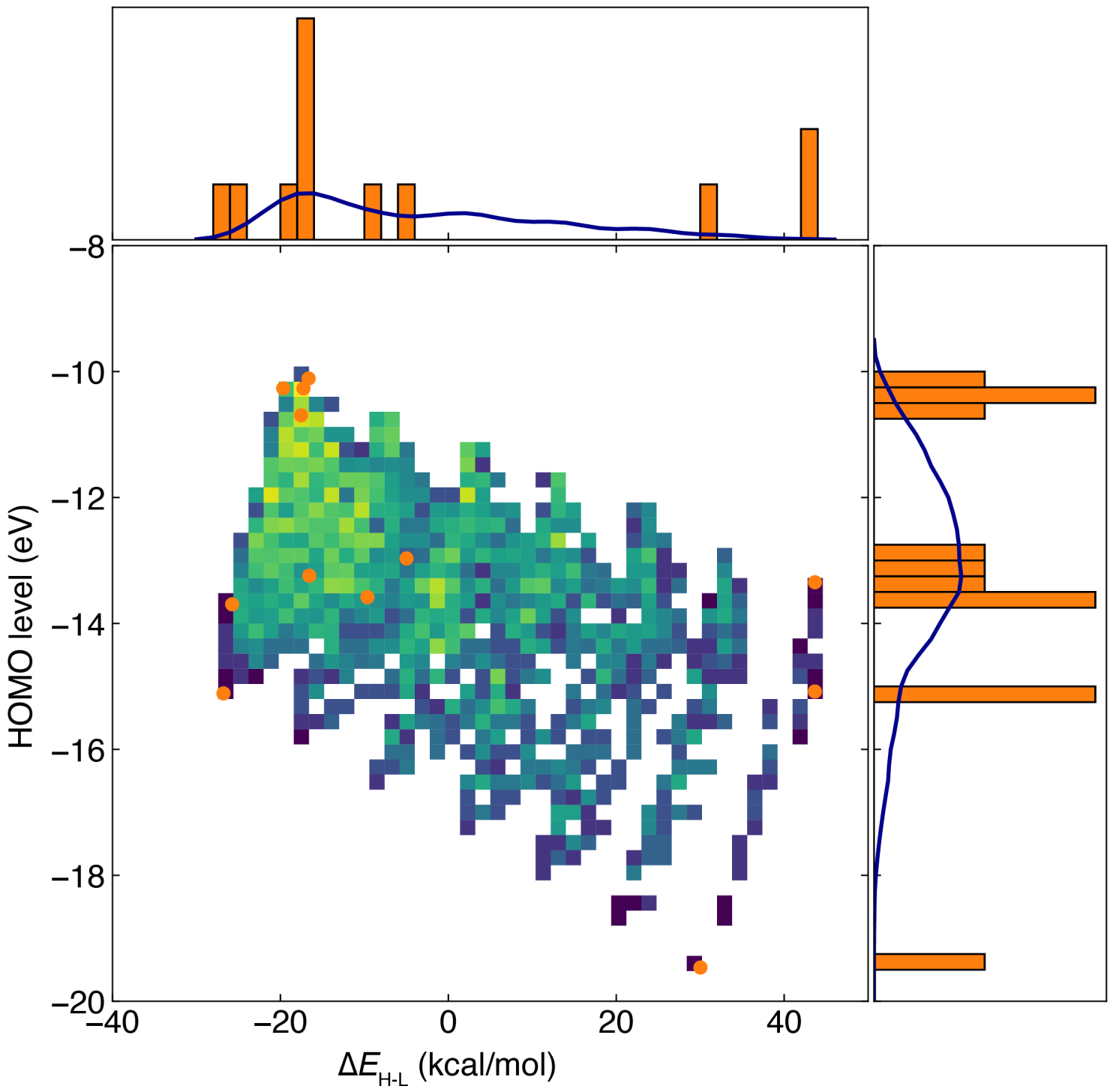

**Figure S6.** Homoleptic-only interpolation for Fe(II) complexes: the $\Delta E_{\text{H-L}}$ (in kcal/mol) vs. singlet HOMO level (in eV) for 12 homoleptic Fe(II) complexes shown as both a scatter plot (middle, orange) and with normalized marginal 1D histograms at top and right (orange bars), respectively. The interpolated values are shown as a 2D histogram colored from low (purple) to high (yellow) density, and the same data is shown as a kernel density estimate on the histogram panes.

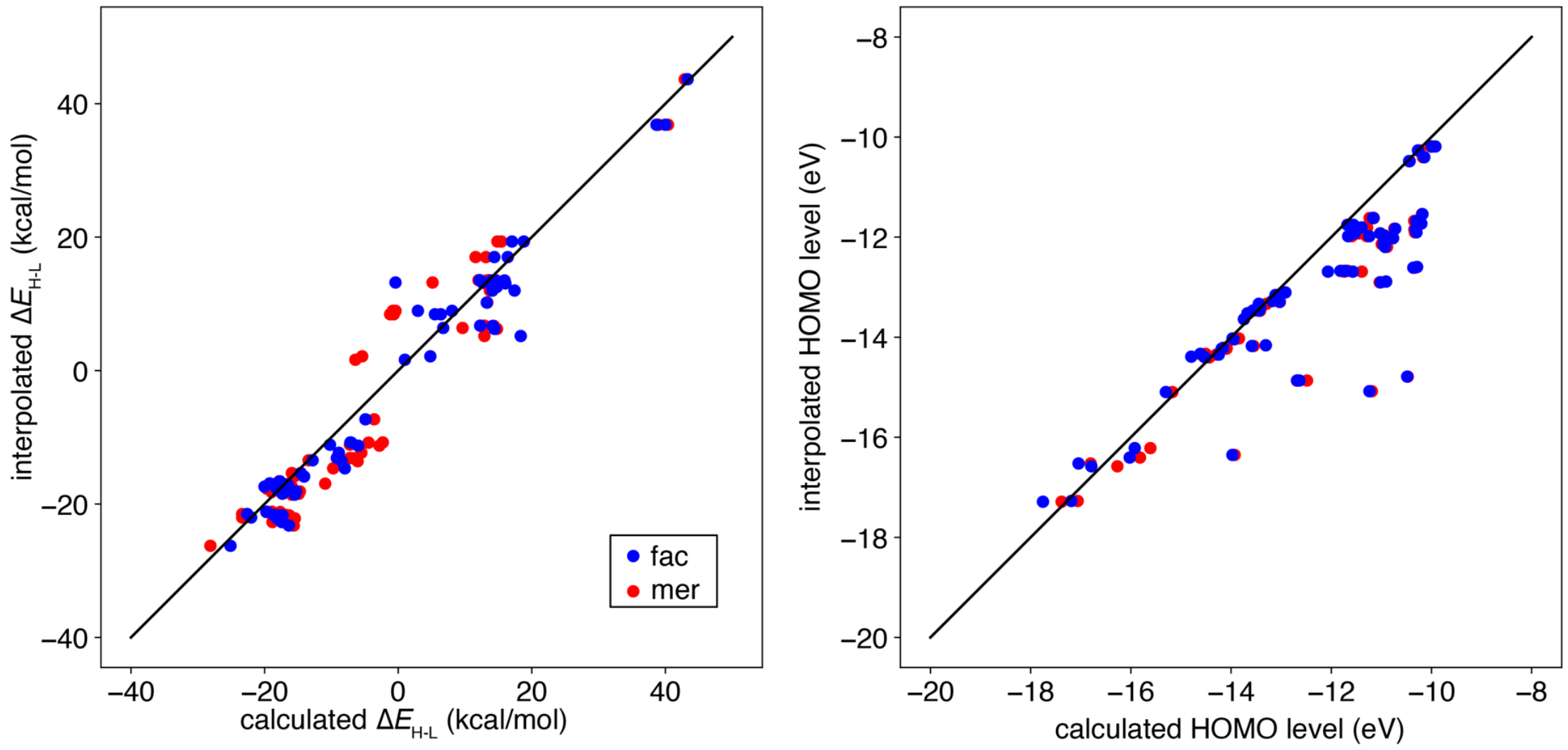

**Figure S7.** Parity plot of calculated vs. HO-only interpolated $\Delta E_{\text{H-L}}$ (in kcal/mol, left) and HOMO level (in eV, right) for 66 *fac* (blue circles) and *mer* (red circles) derived from 12 homoleptic Fe(II) complexes. A black solid parity line is also shown.

**Table S19.** MAEs of homoleptic-only interpolation over the FS/MS 132 complex subset with respect to calculated values.

| | spin (kcal/mol) | HOMO (eV) |
|---|---|---|
| MS | 3.88 | 0.78 |
| FS | 4.02 | 0.77 |

**Table S20.** Binary complexes with their chemical formula and symmetry along with $\Delta E_{H\text{-}L}$ (in kcal/mol) and HOMO level (in eV) obtained through three approaches: homoleptic-only interpolation, FS/MS-augmented interpolation, and explicit calculation. For the cases that are FS or MS complexes, FS/MS-augmented interpolation refers to the explicitly calculated property. As a result, the reported MAEs are obtained over 5 complexes for FS/MS-interpolation but 8 complexes for HO-interpolation. The resulting calculated properties that are in the targeted zone are shown in bold.

| Complex | Sym. | HO-interpolation | | FS/MS-interpolation | | Calculation | |
|---|---|---|---|---|---|---|---|
| | | spin (kcal/mol) | HOMO (eV) | spin (kcal/mol) | HOMO (eV) | spin (kcal/mol) | HOMO (eV) |
| Fe(II)$(MeCN)_5$(CO) | 5+1 | 0.9 | -14.05 | 1.6 | -13.95 | **1.5** | **-13.84** |
| Fe(II)$(MeCN)_3(NH_3)_3$ | MS | -7.3 | -13.28 | -3.6 | -13.17 | **-3.6** | **-13.17** |
| Fe(II)$(CH_3CN)_3(MeOH)_3$ | FS | 9.0 | -13.52 | 2.9 | -13.67 | **2.9** | **-13.67** |
| Fe(II)$(CH_3CN)_3(MeOH)_3$ | MS | 9.0 | -13.52 | -0.8 | -13.65 | **-0.8** | **-13.65** |
| Fe(II)$(ClPyz)_4(CO)_2$ | CS | -1.0 | -15.32 | 2.7 | -13.73 | 4.9 | **-13.63** |
| Fe(II)$(ClPyz)_4(CO)_2$ | TS | -1.0 | -15.32 | 3.1 | -13.70 | 5.7 | **-13.62** |
| Fe(II)$(MeOH)_4(CH_3CN)_2$ | CS | -2.6 | -13.58 | -6.6 | -13.56 | -9.1 | **-13.80** |
| Fe(II)$(MeOH)_4(CH_3CN)_2$ | TS | -2.6 | -13.58 | -9.1 | -13.66 | -13.9 | **-13.67** |
| **MAE** | | 6.3 | 0.54 | 2.5 | 0.11 | | |

**Table S21.** Ternary complexes with their chemical formula and symmetry along with $\Delta E_{H\text{-}L}$ (in kcal/mol) and HOMO level (in eV) obtained through three approaches: homoleptic-only interpolation, FS/MS-augmented interpolation, and explicit calculation. The resulting calculated properties that are in the targeted zone are shown in bold.

| Complex | Sym | HO-interpolation | | FS/MS-interpolation | | Calculation | |
|---|---|---|---|---|---|---|---|
| | | spin (kcal/mol) | HOMO (eV) | spin (kcal/mol) | HOMO (eV) | spin (kcal/mol) | HOMO (eV) |
| Fe(II)$(ClPyz)_4(CNH)_1(NH_3)_1$ | CA | -5.4 | -13.60 | -3.8 | -13.32 | **-3.8** | **-13.37** |
| Fe(II)$(DMTF)_3(CO)_2(H_2O)_1$ | FA | -3.1 | -14.14 | -3.0 | -13.20 | -8.2 | -12.55 |
| Fe(II)$(NH_3)_4(DMTF)_1(CO)_1$ | TA | -4.3 | -14.01 | -3.2 | -13.94 | **-0.2** | **-13.40** |
| Fe(II)$(ClPyz)_4(MeCN)_1(CO)_1$ | CA | -6.9 | -14.23 | -3.8 | -13.38 | **-0.5** | **-13.33** |
| Fe(II)$(MeCN)_4(ClPyz)_1(NH_3)_1$ | TA | -7.7 | -13.12 | -3.6 | -13.03 | **-2.7** | -12.83 |
| Fe(II)$(ClPyz)_3(CO)_2(DMTF)_1$ | FA | -9.0 | -13.29 | -0.1 | -13.96 | **-2.2** | **-13.20** |
| **MAE** | | **4.8** | **0.62** | **2.5** | **0.37** | | |

**References**

1. C. Duan, J. P. Janet, F. Liu, A. Nandy and H. J. Kulik, J. Chem. Theory Comput. **15** (4), 2331-2345 (2019).